\def\BibTeX{{\rm B\kern-.05em{\sc i\kern-.025em b}\kern-.08em
    T\kern-.1667em\lower.7ex\hbox{E}\kern-.125emX}}
\documentclass[conference]{IEEEtran}
\IEEEoverridecommandlockouts

\usepackage{amsmath,amsfonts}
\usepackage{amssymb}
\usepackage{comment}
\usepackage{weiwAlgorithm}
\usepackage{weiwMath}
\usepackage{graphicx}
\usepackage{xcolor}
\usepackage{float}
\usepackage{url}
\usepackage{pifont}
 \usepackage{booktabs}
\usepackage[update,prepend]{epstopdf}
\usepackage{bbding}
\usepackage{tabularx}
\usepackage{multirow}
\usepackage{ifsym}
\usepackage{paralist}
\usepackage{color}

\newcommand{\concat}{\circ}

\newcommand{\eat}[1]{}

\usepackage[scriptsize,tight]{subfigure}

\newcommand{\mydef}[2]{\textit{\textbf{Definition #1} #2.}}

\newcommand{\dis}{\mathit{Dist}}
\newcommand{\seg}{\mathit{Seg}}
\newcommand{\deli}{\mathit{Deli}}
\newcommand{\una}{\mathit{Una}}

\usepackage[all=normal, floats, bibnotes, wordspacing, charwidths, indent, lists]{savetrees}

\makeatletter \skip\footins 8.1pt plus 4pt minus 2pt \floatsep 5pt plus 2pt minus 2pt        \intextsep 5pt plus 2pt minus 2pt       \dblfloatsep 5pt plus 0pt minus 0pt     \dbltextfloatsep 5pt plus 0pt minus 0pt  

\begin{document}

\title{Automatic String Data Validation with Pattern Discovery} 

\newcommand{\szu}{$^{1}$}
\newcommand{\szue}{$^{1 \Letter}$}
\newcommand{\ant}{$^{2}$}
\newcommand{\osaka}{$^{3}$}
\newcommand{\osanag}{$^{3,4}$}
\newcommand{\nagoya}{$^{4}$}
\newcommand{\hkustgz}{$^{5}$}
\newcommand{\uwaterloo}{$^{6}$}

\author{\IEEEauthorblockN{
\szu Xinwei Lin, 
\szu Jing Zhao,
\ant Peng Di,
\osanag Chuan Xiao, 
\szu Rui Mao, 
\szu Yan Ji,
\osaka Makoto Onizuka, \\
\hkustgz Zishuo Ding, 
\uwaterloo Weiyi Shang, 
\szu Jianbin Qin\textsuperscript{\Envelope}
}
\IEEEauthorblockA{
\szu Shenzhen Institute of Computing Sciences, Shenzhen University; \ant Ant Group;  \osanag Osaka University; Nagoya University; \\ \hkustgz The Hong Kong University of Science and Technology (Guangzhou); \uwaterloo University of Waterloo
}

\IEEEauthorblockA{
Email: \{linxinwei2020, zhangxt, mao,jy197541, qinjianbin\}@szu.edu.cn; dipeng.dp@antgroup.com;\\ \{onizuka, chuanx\}@ist.osaka-u.ac.jp; zishuoding@hkust-gz.edu.cn; wshang@uwaterloo.ca}
}

\maketitle

\begin{abstract}

In enterprise data pipelines, data insertions occur periodically and may impact downstream services if data quality issues are not addressed. Typically, such problems can be investigated and fixed by on-call engineers, but locating the cause of such problems and fixing errors are often time-consuming. Therefore, automatic data validation is a better solution to defend the system and downstream services by enabling early detection of errors and providing detailed error messages for quick resolution.
This paper proposes a self-validate data management system with automatic pattern discovery techniques to verify the correctness of semi-structural string data in enterprise data pipelines. Our solution extracts patterns from historical data and detects erroneous incoming data in a top-down fashion. High-level information of historical data is analyzed to discover the format skeleton of correct values. Fine-grained semantic patterns are then extracted to strike a balance between generalization and specification of the discovered pattern, thus covering as many correct values as possible while avoiding over-fitting. To tackle cold start and rapid data growth, we propose an incremental update strategy and example generalization strategy. 
Experiments on large-scale industrial and public datasets demonstrate the effectiveness and efficiency of our method compared to alternative solutions. Furthermore, a case study on an industrial platform (Ant Group Inc.) with thousands of applications shows that our system captures meaningful data patterns in daily operations and helps engineers quickly identify errors.

\end{abstract}


\section{Introduction}

Software system failures inevitably occur in production environments, and diagnosing them is highly challenging. Researchers have proposed numerous techniques for fault localization, automated debugging and repair of the program code~\cite{Boehme2017, Wong2016, Goues2012, Wen2018}. Nevertheless, it is unlikely that all failures are caused by software bugs; the faulty input data can also disrupt system operations, especially for data-intensive scalable computing (DISC) systems (e.g., Apache Spark)~\cite{Zhang2020, Gulzar2021, Teoh2019, Kirschner2020, Zeller2002, Boehme2017}. Recent studies~\cite{Kirschner2020} have highlighted the prevalence of invalid input files in practice, often attributed to missing or non-conforming elements. For instance, research~\cite{DBLP:journals/jss/WangCZW22} has shown that 11\% of issues reported on platforms like Stack Overflow are related to data input and output (IO), underscoring the importance of ensuring data integrity.

On the other hand, data evolves continuously and periodic data insertions are common in enterprise data pipelines (e.g. SSIS~\cite{SSIS}, PowerBI~\cite{PowerBI}) and data 
warehouses~\cite{kozielski2009new}, 
and data warehouses require frequent updates~\cite{kozielski2009new}. However, 
a common data management issue of such updates is that they may introduce errors and compromise data quality and security. These errors may stem from human errors or software bugs and are often difficult to detect and debug, which further poses difficulty in software failure diagnosis. Hence, data quality assurance has become crucial, necessitating effective data cleaning and validation methods. However, addressing data quality issues can be challenging and resource-intensive. Identifying and resolving errors requires substantial human effort, potentially leading to increased costs and reduced efficiency in data-driven systems~\cite{song2021auto}.

\textbf{Data cleaning and error detection.} 
{\color{black}
The primary objective of data cleaning is to address incompatible, dirty data in a dataset. This process includes error detection and repair. Error detection identifies data values that deviate from the ground truth. Existing commercial systems, such as Microsoft Excel~\cite{Excel}, Trifacta~\cite{Trifacta}, and OpenRefine~\cite{OpenRefine}, offer basic error detection features that can help users identify potential quality issues based on predefined patterns.
Several studies have focused on error detection in multi-column tables, often using functional dependencies or denial constraint methods. These methods leverage relationships between columns in multi-column tables to detect errors effectively. Qahtan~\cite{qahtan2020pattern} et.al proposes the concept of pattern function dependency (PFD), which uses patterns to discover correlations among column content in relational data for data cleaning.
However, these methods are typically more suitable for static datasets and may be less effective in environments where data and constraints change frequently. Implementing these solutions in such environments can be costly.}

\begin{figure*}[h]
    \centering  
    \includegraphics[width=0.95\textwidth]{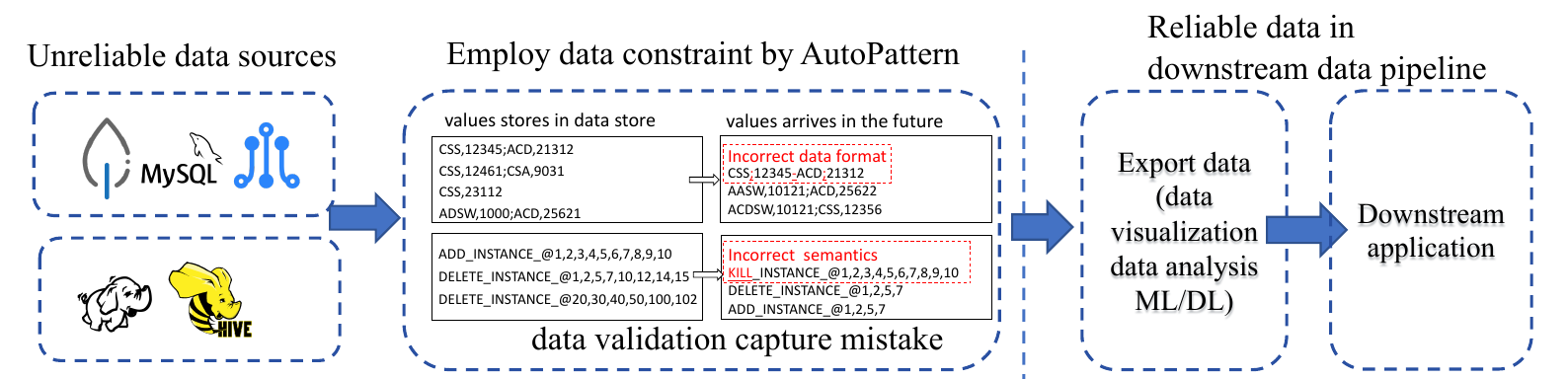}
    \caption{Data pipelines with data validation}
    \label{Fig.data_pipeline}
\end{figure*}

\textbf{Data validation by descriptive constrains and inferred constraints. }
{\color{black} 
The primary goal of a data validation system is to detect errors in data pipelines as early as possible and provide engineers with as much information as possible to fix those errors. Due to the delay caused by changes in the data pipeline, upstream errors are often not discovered until downstream usage of the upstream data. In response to these issues, data validation tools have been proposed. These methods include~\cite{breck2019data,schelter2018automating}, which develops a domain-specific language (DSL) for users to write declarative data constraints. However, in real-world scenarios, manual writing of descriptive constraints can be time-consuming and labor-intensive, particularly for large amounts of data. Therefore, it is necessary to employ automatic data validation techniques to protect the system and downstream services. Some method use inferred constrains to verify data automatically. Auto-validate~\cite{song2021auto} uses inferred regular expression-like patterns to validate string data. And some existing data profiles methods~\cite{padhi2018flashprofile,wang2016fidex,raman2001potter} can automatically generate patterns from the data, which can also be used for data validation. The limitation of these approaches is that the possibility of nested data and the structure of data is not considered. }

{\color{black} 
\textbf{Structural extraction for semi-structural string data. }
When evaluated on production data in an enterprise platform, we have discovered that semi-structured strings account for more than 70\% of production environments, and existing methods~\cite{biessmann2021automated,schelter2019unit} result in a high number of false alarms on semi-structured string data. This illustrates that semi-structured string data is a vulnerable link in the production environment, and is more susceptible to errors and a lack of protective mechanisms.

In semi-structured data, values are often represented as strings, and the structural information is not explicitly expressed. To date, there has been no previous work on applying patterns to data validation for semi-structural string data. Related works for semi-structured data include structure extraction and text segmentation. Structure extraction involves transforming unstructured or semi-structured text or data into a structured form. Notable approaches in this area include Judie~\cite{agichtein2004mining}, CRAM~\cite{cortez2011joint}, ListExtract~\cite{elmeleegy2009harvesting}, and TEGRA~\cite{chu2015tegra}. PADS~\cite{fisher2005pads} focuses on \textit{ad hoc} data sources, which refer to any semi-structured data. These methods define a type description language and a series of tools that can transform semi-structured data. LearnPADS~\cite{fisher2011pads} proposes an automatic method that can learn type descriptions from data without human intervention. FlashFill~\cite{gulwani2011automating} describes a restricted string expression language and proposes an algorithm that can automatically generate a string processing program according to given input-output examples.
However, these structure extraction methods usually do not generate patterns that can be directly used for data validation, and they rely on information at the structural level for data validation scenarios, which may not be accurate enough to capture character-level errors. The significant difference between our work and structure extraction is that we are not only concerned with the results of structure transformation, but also with the impact of structure on data quality.}

{\color{black}
\textbf{Automatic data validation for semi-structural data by inferred patterns. } As shown in the Fig~\ref{Fig.data_pipeline}, the enterprise's data pipeline ingests data from different data sources, and two types of errors may occur in the data.

\textbf{Example 1.1 (incorrect data format)}. In the first part of Fig- ure 1, all historical values of first part follow the same format: several letters (e.g., “CSS”), a comma, and several numbers (e.g., 12345), and the above information may appear recurrently, separated by a de- limiter “;”. An example of erroneous data with incorrect delimiters is shown on the right side. For downstream applications, storage may become inaccessible due to such erroneous value.

\textbf{Example 1.2 (incorrect semantics)}. The second part of Fig- ure 1 shows an example of character-level inconsistency for values. The correct values need to start with “DELETE” or
“ADD”, while the erroneous example starts with “KILL”. This example shows that the validation system should not only extract data format, as shown in Example 1.1, but also be able to mine fine-grained semantics to detect incorrect string values.

\subsection{Our approach and contributions}

\textbf{Our approach. }In this paper, we propose a two-step pattern discovery method for an automatic data validation system. Our solution comprises two steps. 
First, in the skeleton extraction step, we analyze the high-level information of historical data to discover its structure. Second, in the fine-grained semantics extraction step, we use entropy-based character-level generalization and string segmentation to extract fine-grained semantics. We design a cost function to control the trade-off between the generalization and specification of patterns. Additionally, the validation system provides optimization methods such as data augmentation and pattern update strategies to handle the cold start case and data growth.

\begin{figure}[t]
    \centering  

    \includegraphics[width=0.49\textwidth]{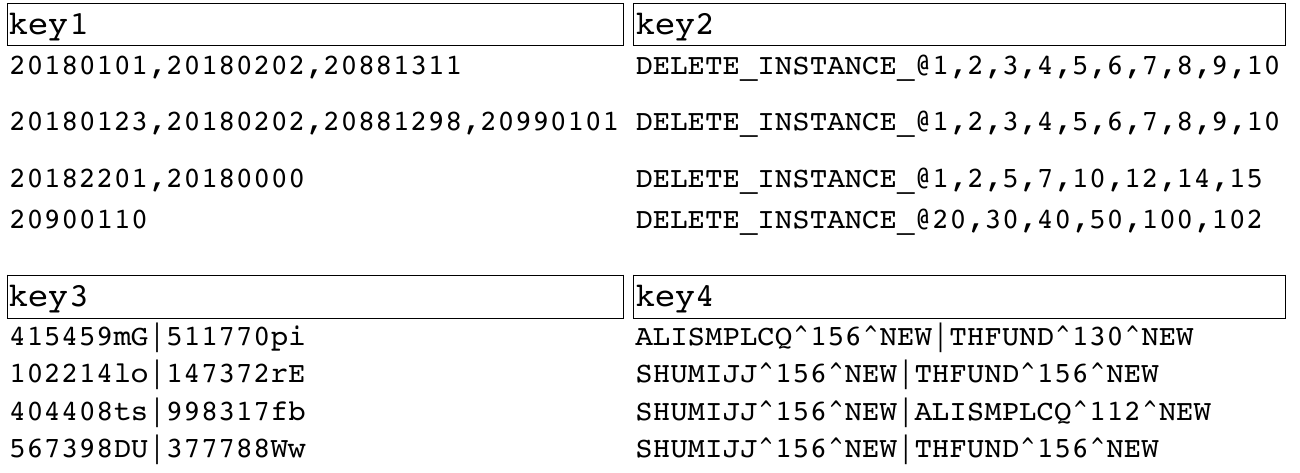}
    \caption{Complex nested data examples.}
    \label{Fig.example}
\end{figure}

\textbf{Contributions.}
In this paper, 
\begin{inparaenum}[(1)]
    \item we propose an efficient automatic data validation system that utilizes automatic pattern discovery techniques to detect erroneous semi-structural string data and address the aforementioned error types.
    \item Our system follows a two-step approach, where we first identify the structural constraints of correct data, followed by mining fine-grained semantics using entropy-based cost functions and character-level generalization.
    \item To evaluate the system's effectiveness and efficiency, we conducted experiments on public and industrial datasets. 
    Our system achieved an average precision of 0.91 and recall of 0.89, outperforming alternative solutions. Additionally, we conducted a case study on a industrial platform (Ant Group Inc.) with thousands of applications, 
    demonstrating that data validation improves data quality and provides more accurate information for error location and resolution by engineers.
\end{inparaenum} }

\textbf{Paper outline.}
The rest of this paper is organized as follows.
\begin{inparaenum} [(1)]
\item In the Section 2, we define the data validation problem and our self-validate framework. 
\item In the Section 3, we define our novel domain-specific-language (DSL).
\item In the Section 4, we present the skeleton extraction algorithm.
\item In the Section 5, we present the fine-semantics extraction algorithm.
\item In the Sections 6, we present the data augment and pattern update method. 
\item In teh Section 7, we empirically evaluate our solution in severals datasets and report their results. 
\end{inparaenum}
 \section{Preliminaries and Framework}
In this section, we define some definitions and introduce
the framework of our data validation system. 

\subsection{Terminology}

\noindent{}\mydef{2.1}{Atom pattern}
Atom pattern is a series of predefined or user-defined atom constraints over strings. 
The atom pattern $\alpha:\text{String} \rightarrow \text{Int}$ is a function, which gives a string s, and returns the length of the longest prefix of s that satisfies the pattern constraint. $\alpha(s)=0$ indicates match failure of $\alpha$ on s. The alias of $\alpha(s)$ is also called the pattern matching length.

\noindent{}\mydef{2.2}{Pattern language} 
A pattern is an arbitrary sequence of atom patterns,
each containing low-level logic for matching a sequence
of characters. A pattern P describes a string s, $i.e. P(s)=\textrm{True}$, if the atoms in P match contiguous non-empty substrings of s, ultimately matching s in its entirety.

\noindent{}\mydef{2.3}{Data profile}
Give a set of strings, learn a pattern P such that $\forall s \in S: P(s)=\text{True}$, the goal of the data profile is to learn a set of patterns that properly summarize the given dataset.

Intuitively, it is difficult for a data profile program to choose what patterns are suitable for data validation if it only looks at historical data (It may be too general or too specific). However, we as humans know what patterns are suitable base on human's high-level insight for data.

\noindent{}\mydef{2.4}{Data validation}
Given origin dataset $S_{0}$, a data profile pattern $P_{0}$ and human's optional feedback $F_{0}$, data validation algorithm fine-tunes the pattern $P_{0} \stackrel{F_{0}}{\rightarrow} P_{0}^{'}$ 
And use the Pattern $P_{0}^{'}$ to validation the incoming data $\bigtriangleup S_{1}$. So Data validation is a dynamic process. i.e.  $P_{0} \stackrel{F_{0}}{\rightarrow} P_{0}^{'},P_{0}^{'}\stackrel{F_{1}}{\rightarrow} P_{1}^{'}.....$
The main purpose of our work is to compose constrain for incoming data so that the quality of the dataset can be proved.

\subsection{Solution framework}
\begin{figure}[htbp]
\centerline{\includegraphics[scale=0.4]{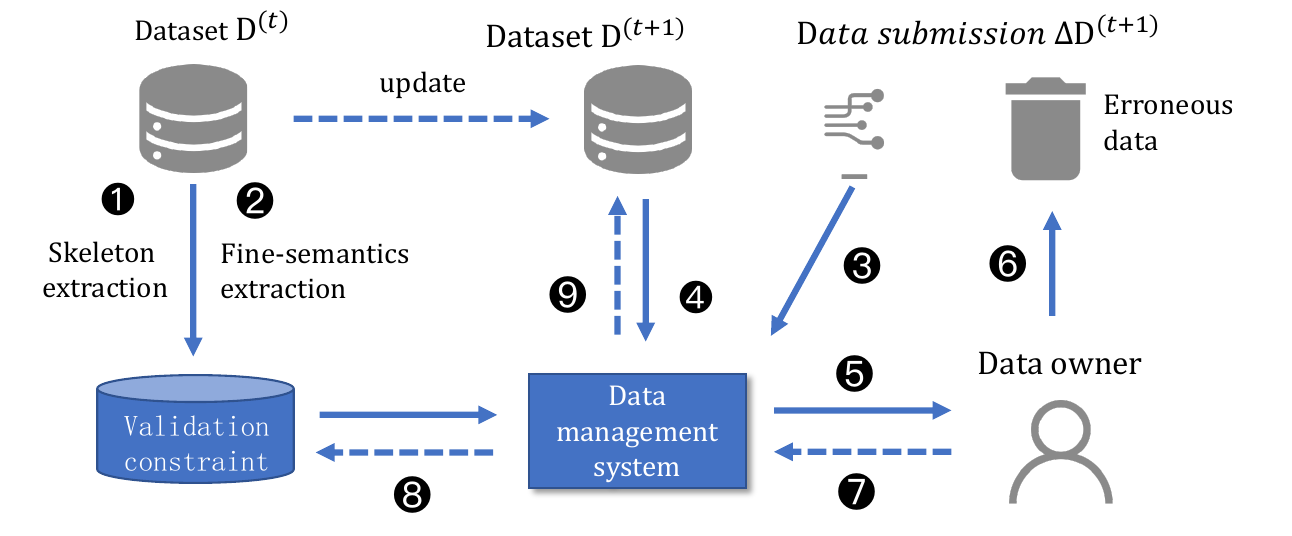}}
\caption{Self-validate data management system framework.}
\label{Fig.framework}
\end{figure}

Fig.~\ref{Fig.framework} shows the framework of our self-validate data 
management system. Given a snapshot of a dataset $D^{(t)}$, the self-validate 
data management system generates a skeleton $H_(D^{(t)})$ by 
skeleton extraction (\ding{202}) and refines it to $P_f(D^{(t)})$ 
by fine-grained semantics extraction (\ding{203}). With the submitted data 
$\Delta D^{(t+1)}$ at time $t+1$, the system checks whether 
$\Delta D^{(t+1)}$ is compatible with $P_f(D^{(t)})$ 
(\ding{204}). 
If the system confirms $\Delta D^{(t+1)}$ is compatible with  
$P_f(D^{(t)})$, $\Delta D^{(t+1)}$ passes the validation (\ding{205}) 
and the system updates the dataset with the submitted data. 
Otherwise, the system regards the submitted data as erroneous and asks the user 
who has submitted the data to reconfirm if the submitted data is correct (\ding{206}). 
If the user reports that the data is incorrect, the system rejects the data (\ding{207}). 
If the user insists that the data is correct (\ding{208}), the system updates the 
constraints (\ding{209}) and the dataset with the submitted data (\ding{210}). To put 
it succinctly, our system uses historical data to generate appropriate constraints and 
uses them to validate future data. 

 \section{Domain specific language}
\begin{figure}[t]
    \centerline{\includegraphics[scale=0.35]{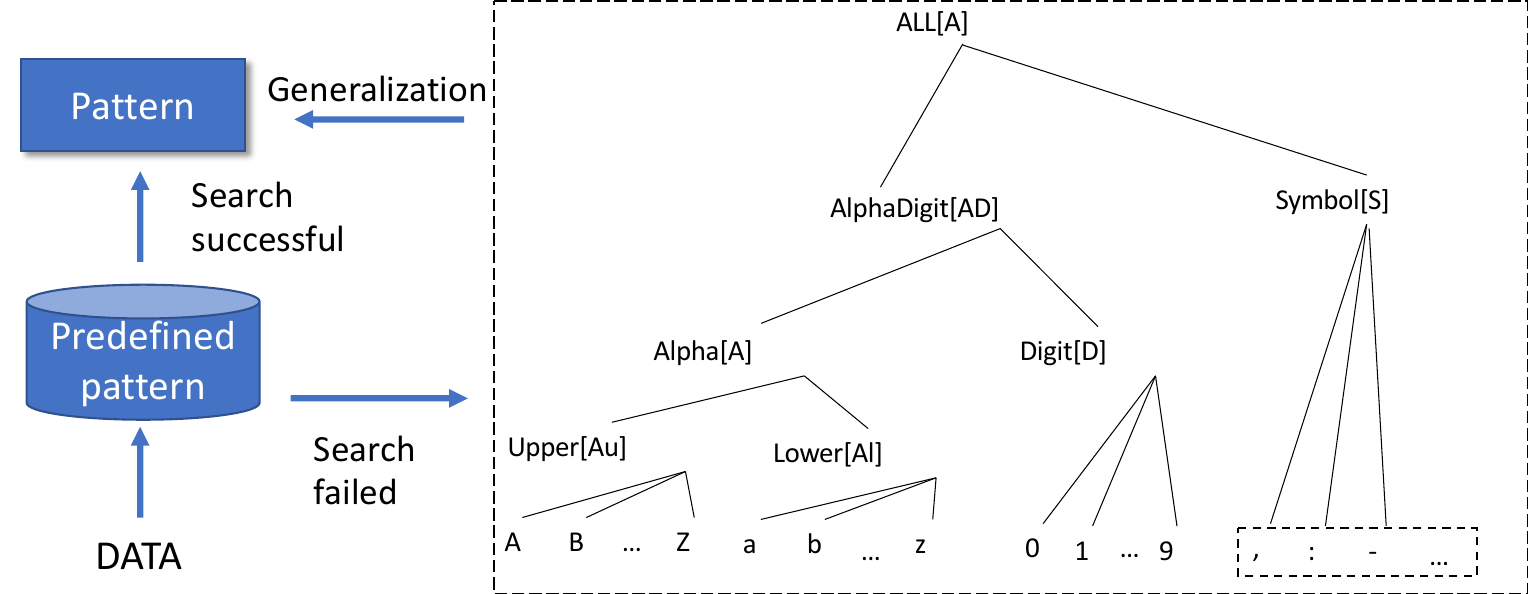}}
    \caption{Generalization tree}
    \label{Fig.tree}
    \end{figure}

Now we present a novel domain-specific language (DSL) for patterns and define a 
specification over a given set of strings. Our DSL allows users to easily augment it with the new atom. (A standard predefined generalization tree~\cite{padhi2018flashprofile,song2021auto}) is shown in Fig.~\ref{Fig.tree}. And we prose a novel two-step method to learn the consistent pattern for data validation.

\noindent{}\mydef{3.1}{The generalization tree}
We deploy a generalization tree where leaf 
nodes represent single characters and internal nodes represent the atom pattern.
The mapping function $m: \textrm{character} \to \textrm{atom pattern}$ indicates the 
generalization of character base generalization tree $G$. The generalization tree can be arbitrarily defined in combination with the actual situation of the application.

\noindent{}\mydef{3.2}{Patter match}
A pattern is a sequence of atoms. We use $\concat$ as concatenation operator. The sequence $p = \alpha_{1} \concat \alpha_{2} \concat ... \concat \alpha_{k}$ of atoms
defines a pattern. We define the pattern match as a function: $\textrm{String} \rightarrow \textrm{Int}$. The function returns the longest prefix match length of $p$ in string $s$.
\begin{equation}
    p(\textrm{s})= \alpha_{1}(s[: \textrm{prefix}]) + (\alpha_{2} \concat \alpha_{3} ... \alpha_{n} (\textrm{s[prefix:]}))
\end{equation}

where given string $s$, the match length function return the longest match length of  $\alpha_{1}$ in string s and plus the match length of  $p = \alpha_{2} \concat \alpha_{3} ... \alpha_{n}$ in matching the suffix of string $s[:\textrm{suffix}]$. If a part of the prefix fails to match, the match ends. $p(s)=0$ indicates match failure of pattern $p$ on s. $p(s)=len(s)$ indicates pattern match all string. 
We adopt the idea of greedy matching to avoid additional complexity.

\noindent{}\mydef{3.3}{Base type and structural definition}
\begin{figure}[h]
    \centerline{\includegraphics[scale=0.45]{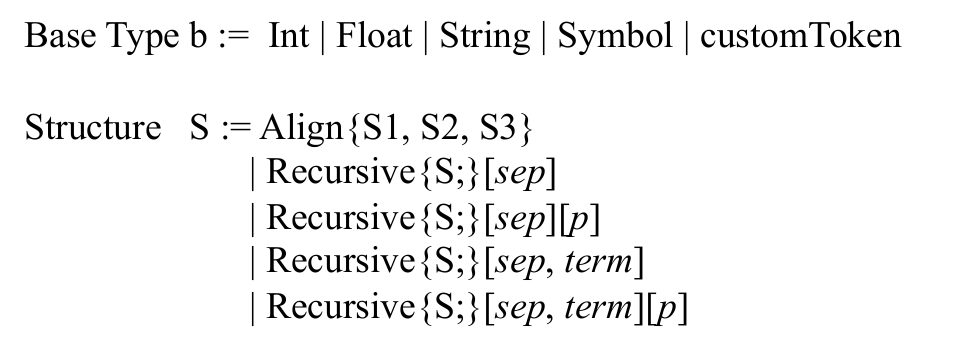}}
    \caption{Base type and structural definition}
    \label{Fig.basetype}
  \end{figure}

{\color{black}
The \textit{base type} is generally int, float, string and other custom's defined tokens. On the structural definition. \textrm{Align\{$S_{1},S_{2}...S_{k}$\}}, which means that the data should contain a sequence of structures with $S_{1}, S_{2}... S_{k}$ are matched one by one. And the \textrm{Recursive\{$S;$\}[$sep$]} means that the data is repeated any number of times in the structure S
, and $sep$ as separators. $p$ means the structure repeat $p$ times. }

\noindent{}\mydef{3.4}{Syntax and semantics of pattern}

\begin{figure}[h]
    \centerline{\includegraphics[scale=0.45]{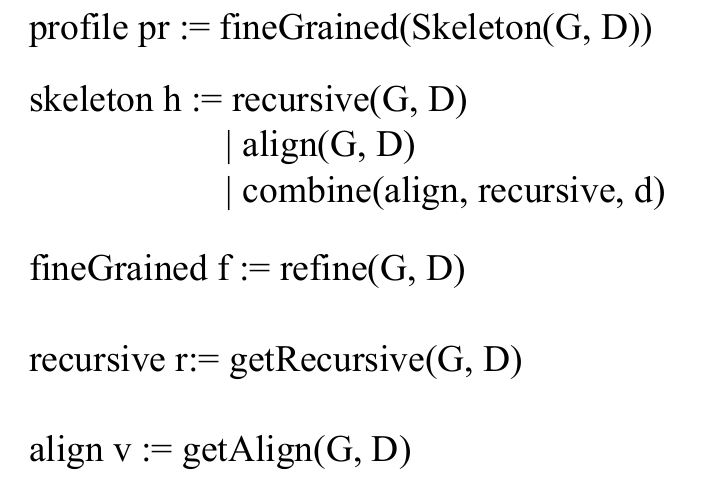}}
    \caption{Syntax and semantics of pattern}
    \label{Fig.dsl}
  \end{figure}
where $G$ denotes a predefined generalization tree, $D$ denotes a list of strings.
The proposed DSL provides predicates for skeleton extraction and fine-grained extraction, which is significantly different from the previous approaches.
We divide the skeleton in the data into two categories. The first is \textit{Recursive}, which means that the DSL is able to capture the characteristics of repeated cycles in the data. The second is \textit{Align}, which means the DSL is able to capture the character of align data. By combining (i.e. DSL operator \textit{combine}) the two skeleton types, the algorithm can effectively discover the skeleton in the data, we use parameter $d$ to control the
nesting depth of the skeleton to avoid exponential complexity. In section 4, we propose a skeleton extraction algorithm to mine the optimal skeleton (i.e. DSL operators \textit{getRecursive} and \textit{getAlign}). In section 5, we propose a fine-grained extraction algorithm base on the skeleton (i.e. DSL operator \textit{refine}).

\noindent{}\mydef{3.5}{The pattern-based distance}
A widely used string similarity measure is Levenshtein distance (a.k.a. edit distance), 
which measures the minimum number of edit operations (insertion, deletion, 
and substitution of a character) that transforms one string into another. However, simply 
using edit distance for our problem often fails to evaluate the semantic (dis)similarity 
of data because it does not constrain underlying patterns. 
We define the pattern-based distance between strings $x$ and $y$ 
as the minimum distance incurred by patterns.

For instance, 
given two strings 
``\textrm{ID:1234-5678-9012-3456}'' and ``\textrm{ID:abcd-abcd-ab+d-0adb}'', the first 
one is correct, while the second one is incorrect for having a ``+'' in the third group. 
Suppose edit distance is used to compute the similarity between groups. In the first 
string, each pair of group yields an edit distance of 4, while in the second string, the 
incorrect group ``\textrm{ab+d}'' only yield an edit distance of 1 to its correct group 
``\textrm{abcd}''. Edit distance fails in this example because the incorrect configuration 
results in a smaller distance, which is counterintuitive. 

Based on the intuition that strings with the same underlying pattern should be more similar. We define a pattern-based distance based on a given generalization tree. As shown in algorithm~\ref{alg:get pattern-distance}, if the strings x and y are the same, then the distance between x and y is 0. We
predefine the cost of mapping character to atom pattern named generation cost(for short GC), and we use the predefined GC to quantize the distance between the string and general pattern. When the strings x and y are mapped to an atom pattern, x and y are equivalent, and we take the minimum cost of mapping to the equivalent atom pattern as the pattern-based distance between x and y. So we define an algorithm for distance solving based on the idea of solving the nearest common ancestor node.

\begin{algorithm}[t]
    \DontPrintSemicolon
      
      \Input{$x, y$}
      \Output{$distance$}
      \If{$x == y$}{
        \Return{0} 
      }\ElseIf{$x != y$}{
        \textrm{parent} $\gets$ $\textrm{findNearestParent}(x, y, G)$ \;
        \If{parent $==$   \textrm{Null}}{
            \Return{inf}
        }
        distance $\gets$ GC(x, parent) + GC(y, parent) \;
        \Return{distance}
      }
    
     \caption{getPatternBasedDistance(x, y)} \label{alg:get pattern-distance} 
    \end{algorithm}

\section{Skeleton extraction}\label{sec:skeleton}

\noindent{}\mydef{4.1}{Atomic data}
Given a string, we define there is no subdomain inside the string as atomic data. 
This corresponds to the first normal form(1NF) of the relational model.

\noindent{}\mydef{4.2}{Skeleton extraction} Given a dataset $S$, the skeleton extraction invokes learning (1) a partition of $H_{1} \concat H_{2} \concat ... \concat H_{k}$. (2) a sequence of pattern $P_{1} \concat P_{2} \concat ... \concat P_{k}$, where each $P_{i}$ is a regex-like pattern describe the structure of $H_{i}$.
For unstructured data and semi-structured data, the purpose of Skeleton extraction is to find a suitable segmentation of the data. 

\noindent{}\mydef{4.3}{Optimal skeleton extraction} Given a dataset $S$, the maximum limit number k of partition, a pattern-based distance to measure the similarity of two strings. The purpose of Skeleton extraction is to find a suitable segmentation 
for the data. We define an objective function $H_{opt}$, where $d(H_{i})$ denotes the pattern-based distance among the partition $H_{i}$. $d(H_{i})$ computes each pair among the partition and accumulates them together. The goal of the optimization problem is to minimize the global pattern-based distance among each segmentation so that data can be correctly segmented
into atomic data. 

\begin{equation}
    \label{equa: optimal skeleton}
    {H_{opt}}  = \argmin_{\{H_{1}, H_{2}...H_{k}\}  \atop s.t. H=\sqcup_{i=1}^{k}H_{i} }\sum_{i=1}^{k}(d(H_{i})) 
\end{equation}

Combining the types of skeleton that actually exist in the data, we propose two basic splitting methods include recursive splitting and vertical splitting. The combination of these two basic splitting methods can discover the skeleton of nesting that exist in the data.

\textit{Theorem 1. } Optimal skeleton extraction is NP-hard.

\eat{
\textbf{Pattern-based edit distance}
The distance used in the skeleton algorithm should capture the likelihood of value $d_{1}$ and value $d_{2}$ in the same underlying domain. 
To capture the feature of the domain of data and avoid the data sparsity. 
we define pattern-based edit distance to measure the similarity between two values. 
Traditional edit distance $d(a,b)$ is the minimum-weight series of edit operations that transforms a into b.
 In pattern-based edit distance, the origin of different characters can be considered the same after the generalization. }

\subsection{Recursive splitting algorithm}

\noindent{}\mydef{4.4}{Splitting operator} Given a value $l$, $\seg$ and $\deli$ present the segment and delimiter parts of $l$ respectively. Splitting operator $h(l)$ segments $l$ into the form of $\{\seg_{1} \deli_{1} \seg_{2} \deli_{2} ... \seg_{t}\}$. 

\noindent{}\mydef{4.5}{Recursive splitting} Recursive splitting is all possible splitting of $l$, denoted by $H(l):=\{h_{1}(l), h_{2}(l),...,h_{m}(l)\}$. Given a set of values $\mathcal{L}:=\{l_{1},l_{2},...,l_{n}\}$, the recursive splitting of $\mathcal{L}$ is the set of recursive splitting of each $l$ in $\mathcal{L}$, denoted by $H(\mathcal{L})=\{H(l_{1}),H(l_{2}),...,H(l_{n})\}$.

\noindent{}\mydef{4.6}{Recursive distance} Recursive distance is defined as the sum of the pattern-based distance of each segment pair. 
\begin{equation}
    \dis_r(h(l))=\sum_{1\leq i < j \leq t} \dis(\seg_{i}, \seg_{j}).
\end{equation}

We observe that the typical characteristic of recursive structure is that the recursive parts are connected using some special symbol, and each part connected by delimiter is similar. It means given a set of values $\mathcal{L}$ belong to the recursive structure. $\forall l \in \mathcal{L}$ have a recursive splitting $h(l)$. The each segment $\seg_1, \seg_2,...$ is similar. Therefore finding a proper recursive structure can be transformed to find a delimiter that has minimal recursive distance. 

The optimal segmentation of $l_j$ against the segmenting of $h(l_i)$ is as follows.
\begin{equation}\label{eq:hopt}
    h_{opt}(l_j, h(l_i))=\operatorname*{argmin}_{h(l_j) \in H(l_j)} (\dis_r(h(l_i)) + \dis_r(h(l_j))).
\end{equation}

Algorithm~\ref{alg:recursive-split} shows pseudo-code for recursively finding segmentation. In line 1, we define an empty set to store the skeleton
candidate. In line 2, the algorithm gets all possible segmentation in dataset $\mathcal{L}$. From line 3 to line 4, the algorithm enumerates each value in the dataset and each possible segmentation corresponding to the value.  From line 5 to line 8, the algorithm finds the best segmentation for each value $l_{j}$ against the segmentation of $h(l_{i})$ and accumulates the distance among $l_{i}$ and $l_{j}, i \neq j$. Finally, the algorithm gets the optimal skeleton by the skeleton which has minimum skeleton distance. 

\textbf{Time complexity.}
The above algorithm takes $O(m \cdot n^{2})$-time, where $m$ is the size of recursive segmentation of $\mathcal{L}$ and $n$ is the size of input data $\mathcal{L}$. The computation of recursive distance takes $O(k^{2})$-time, where k is the size of the segment of value $l$. And in order to avoid repeated calculations, it can store the corresponding results according to the index.

\begin{algorithm}[t]
\DontPrintSemicolon
  
  \Input{$\mathcal{L}=\{l_{1},l_{2}...l_{n}\}$}
  \Output{$H$}
$C_{set}  \gets \emptyset$ \;
    $H(l_{1}),H(l_{2})...H(l_{n})\gets$ Get all recursive segmentation of $\mathcal{L}$ \;
    \ForEach{$l_{i} \in \mathcal{L}$}{
        \ForEach{$h(l_{i}) \in H(l_{i})$}{
                $\dis_{sum}=0$ \;
                \ForEach{$l_{j} \in \mathcal{L},j \neq i$}{
                Get $h_{opt}(l_j, h(l_i))$ according to Equation~\ref{eq:hopt} \;
                $\dis_{sum} += \dis(h_{opt}(l_j, h(l_i)))$ \;
                }
                $C_{set}$ $\gets$$ C_{set} \cup \textrm{GetSkelton}(r(l_{i}))$ \;
                
            }
        }
        $H$ $\gets$ Get skeleton according to the minimum disatnce from $C_{set}$ \;
    \Return{$H$}

 \caption{Recursive splitting algorithm} \label{alg:recursive-split} 
\end{algorithm}

\subsection{Vertical splitting algorithm}

\noindent{}\mydef{4.7}{Vertical splitting}
Given a set of values $\mathcal{L}=\{l_{1},l_{2}...l_{n}\}$, $minLength(\mathcal{L})$ is the smallest length of $l\in \mathcal{L}$. The vertical splitting operator $v(l)$ segments $l\in \mathcal{L}$ into the form of $\{\seg^l_{1} \deli^l_{1} ... \seg^l_{t} \una^l \}$, where $Length(l) = minLength(\mathcal{L}) + Length(\una^l)$. 
All possible vertical split of $\mathcal{L}$ can be denoted by $V(\mathcal{L})=\{V(l_1), V(l_2)...V(l_n)\}$.

\noindent{}\mydef{4.8}{Vertical distance}
As for an aligned value pair $l_{i}, l_{j}$, vertical distance is as follows.
\begin{equation}
    \begin{split}
        \dis_v(l_{i}, l_{j})&=\sum_{1\leq k\leq t}\dis(\seg^{l_i}_{k},\seg^{l_j}_{k}) \\
        &+ \sum_{1\leq k\le t}\dis(\deli^{l_i}_{k},\deli^{l_j}_{k}).
    \end{split}
\end{equation}
As for an unaligned value pair $l_{i}, l_{j}$, vertical distance is as follows
\begin{equation}
    \dis^{'}_v(l_{i}, l_{j})=\dis_v(l_{i}, l_{j})+\mathcal{C}_{unalign},
\end{equation}  
where $\mathcal{C}_{unalign}$ denotes the compensation of unaligned values.

\begin{algorithm}[t]
    \DontPrintSemicolon
      
        \Input{ $D=\{l_{1},l_{2}...l_{n}\}$, $G$}
        \Output{ $\mathcal{H}$}
        $C_{set}  \gets \emptyset $ \;
        $V(l_1), V(l_2)...V(l_n) \gets$
        Get all transpose vertical segmentation of $D$ \;
        \ForEach{$V(l_i) \in V(D)$}{
            $\dis_{sum}=0$ \;
            \ForEach{$\textrm{verticl} \in V(l_i)$}{
                    $v_{dist} = \dis_v(v(\textrm{verticl}))$ \;
                    $\dis_{sum}+=v_{dist}$ \;
                    $C_{set}$ $\gets$$C_{set} \cup \textrm{GetSkelton}(v(\textrm{verticl}))$ \;
                }
            }
        $H$ $\gets$ Get skeleton according to the minimum disatnce from $C_{set}$ \;
        \Return{$H$}
    
     \caption{Vertical splitting algorithm} \label{alg:vertical-split} 
    \end{algorithm}

The pseudo-code is shown in Algorithm~\ref{alg:vertical-split}.
In the line 1, the algorithm declares an empty set to store candidate skeletons, and in the line 2, the algorithm takes all possible vertical segmentation in the data and transposes them.
In the line 3, the algorithm enumerates the parts after the transformation, that is, enumerates the parts that may belong to the same skeleton between different data in the same dataset, and solves the corresponding vertical distance. In the line 9, the algorithm solves the optimal distance is used to select the optimal structure from the candidate set.

\textbf{Time complexity.}
The above algorithm runs in $O(nm)$-time, where $n$ is the size of vertical segmentation and $m$ is the size of the vertical part in $V(l_{i})$.

\subsection{Get skeletons algorithm}

\begin{figure}[t]
    \centerline{\includegraphics[scale=0.35]{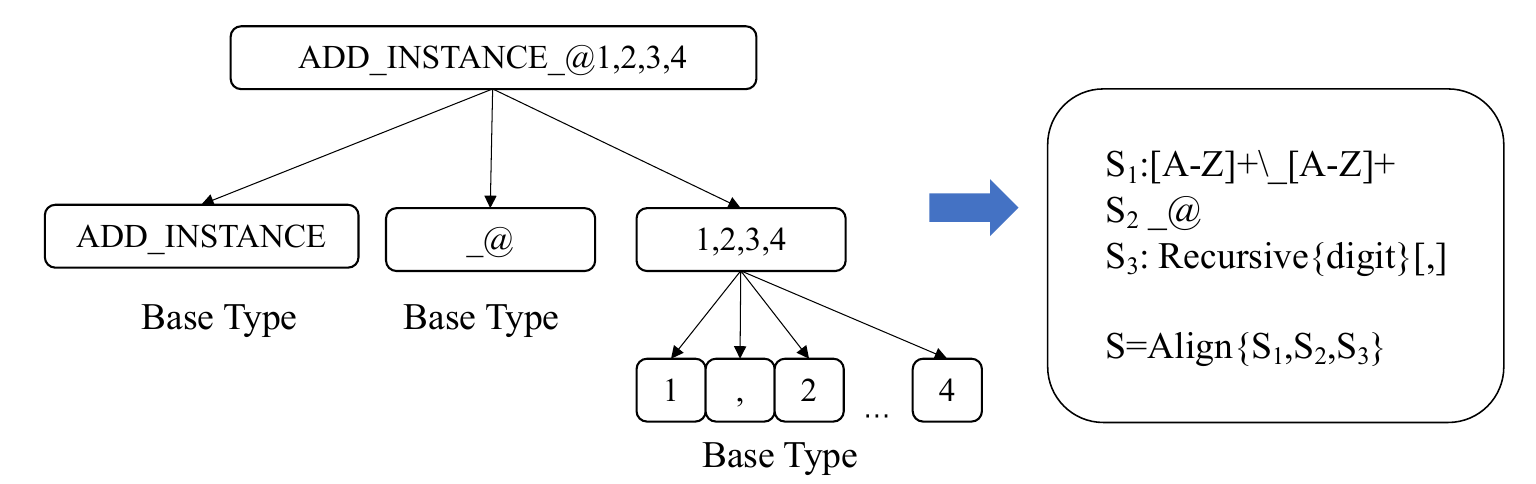}}
    \caption{Skeleton extraction example: given a set of data(only one piece of data is shown here for brevity), the skeleton extraction algorithm uses delimiters to divide the data and then recursively solves it using algorithm~\ref{alg:recursive-split} and algorithm~\ref{alg:vertical-split}. Until the base type is encountered, $S$ splices different sub-structures $S_{1}, S_{2}, S_{3}$ to get the final skeleton.}
    \label{Fig.skeleton}
  \end{figure}
  
  As shown in Fig.~\ref{Fig.example}, the data in the real world is more complicated. 
If we directly consider these complex data, it is difficult to find the skeleton in the data.
However, if we can separately consider the sub-structures after segmenting the complex data,  it is not difficult to find the skeleton of the data. Among them, the difficulty of finding the skeleton lies in: distinguishing the sub-recursive structure inside the data from other structures. For example, as the ``\textrm{Key2}'' shown in Fig.~\ref{Fig.example}, the digit parts ``\textrm{1,2,3,4,5,6}'' in $v_{1}$ belong to recursive structure, the digit may appear recurrently. 
To solve this difficulty, we propose 
a recursion algorithm as shown in Fig~\ref{Fig.skeleton}. The method first enumerates all possible segmentation and judges whether the current sub-structure is already a \textit{base type} that cannot be further segmented.
If it already belongs to \textit{base type}, the algorithm will no longer segment the data. And if the current sub-structures do not belong to the \textit{base type}, the algorithm will divide the current sub-structures again using algorithm~\ref{alg:recursive-split} and algorithm~\ref{alg:vertical-split}. The algorithm will select the current local optimal skeletons according to the pattern-based distance.
The algorithm determines the current segmentation situation with a local optimal idea, and in general, dividing for non-local optimal results usually leads to worse results.

\subsection{Scoring skeletons}
After generating skeletons, there may be multiple skeletons with the same minimum distance. the core challenge in the system is to determine a skeleton the system inferred before is suitable for data validation. Please remember that the system does not require users to provide additional negative examples. Because sometimes it is difficult for users to describe what incorrect data looks likes.
Intuitively, a pattern constraint $P$ extracted from dataset $S$ is a ``good'' constraint, 
if it accurately describes the underlying domain of historical data and data submitted in the future.

We use $Cov({\mathcal{P}})$ to denote the coverage of a generated constraint 
set over historical data. Higher coverage suggests better generated constraints. 
Special symbols usually play an important role 
in the structure of recognizing characters in many scenarios. We define the Symbol-delimiter score, which represents the total coverage of the delimiter symbol in 
the whole skeleton. Intuitively, the more special symbols we consider as 
We sort the candidate skeletons generated in skeletons extraction according to the descending order of coverage, descending order of Symbol-delimiter score, and priority of coverage, and select the top k skeletons patterns.

 \section{Fine-grained semantics extraction}\label{sec:fine-grained}
In this section, we realize that skeleton extraction on data can only generate patterns at the skeleton level. The accuracy of the skeleton-level pattern is not enough for the data validation. Therefore, we use the fine-grained method to refine the pattern on the basis of the skeleton to capture the semantics of the character level. Intuitively, the generalization of data can be viewed as the process of mapping upwards on the generalization tree. And the refinement of the data can be seen as the downward search of the nodes of the generalization tree. Our goal is to find a suitable trade-off between the generalization and specification of the data. This can be seen as the process of traveling on the generalization tree. So we use entropy to describe the generalization of the pattern, use entropy and pattern matching length to define the cost function, and further formally define the fine-grained semantics extraction problem as an optimization problem.

\noindent{}\mydef{5.1}{Entropy}
Information entropy can be used to represent a measure of uncertainty in random variables. The greater the entropy, the greater the amount of information contained in the data.  Fine-grained semantic extraction can be regarded as a process of increasing entropy. So we use entropy to describe the amount of information for the patterns. And 
\begin{equation}
    Ent = -\sum_{n = 1}^{k}P(p_{i})logP(p_{i}) + \beta
\end{equation}

\noindent{}\mydef{5.2}{Cost function}
The cost function is used to select the most proper pattern that profiles the dataset $S$.
For a pattern $P=p_{1} \concat p_{2} \concat ... \concat p_{k}$, we define the cost function
with respect to the given dataset $S$ and the generalization tree $G$. The cost function balances 
the trade-off between generalization and specification. That is because there is a lot of data error caused by spelling error.
The aim of the data validation method is to mine patterns automatically and avoid data errors in submitting data as soon as possible.  The main purpose of the cost function is to control the degree of data refinement so that the pattern after refinement can capture the fine-grained semantics of data.
\begin{equation}
    Cost = \sum_{n = 1}^{k} \alpha(p_{i})Ent(p_{i})
    \label{equa:cost function}
\end{equation}

\noindent{}\mydef{5.3}{Optimal cost function}
Give dataset $S$, the generalization tree $G$, the origin pattern $P$, the purpose
of finding optimal cost function is to find out a proper pattern that can represent the high-level structure information and capture the character level semantics, so it can be used in data validation scenarios.
\begin{equation}
    Optimal \dot{Cost} = \argmin_{p \in P} \sum_{n = 1}^{k} \alpha(p_{i})Ent(p_{i})
\end{equation}

To solve the problem of optimization of the cost function, as shown in Algorithm~\ref{alg:fine_grained} and~\ref{alg:fine_grained_travel}, we employ a greedy-based method. 
From line 3 to line 9 in Algorithm~\ref{alg:fine_grained_travel}, the algorithm chooses the appropriate level of generalization according to the cost function. In line 3, the algorithm enumerate all children nodes of current nodes, which means refining the current pattern base on the generalization tree. From line 5 to line 9, the algorithm compares the origin cost function and refines the cost function to choose whether specific to the current pattern.

\begin{algorithm}[t]
    \DontPrintSemicolon
      
      \Input{$H=\{h_{1},h_{2}..h_{n}\},S=\{s_{1}, s_{2}...s_{n}\}, G$}
      \Output{$P$}
        $P \gets \emptyset$ \;
        \ForEach{$h_{i}, s_{i} \in zip(H, S)$}{ 
            $P_{fine} \gets \emptyset$ \;

            \ForEach{$p_{seq} \in h_{i}, s_{seq} \in s_{i}$ }{
            $p_{fine_{seq}}$ = greedyTravel($p_{seq}, s_{seq}, G$) \;

            }
            $P_{fine} \gets P_{fine} \cup p_{fine_{seq}}$ \; 
            
        }
        $P \gets P \cup P_{fine}$ \; 
        \Return{P}
    
     \caption{Fine-grained semantic extraction} \label{alg:fine_grained} 
\end{algorithm}

\begin{algorithm}[t]
    \DontPrintSemicolon
      
      \Input{$pattern, segment, G$}
      \Output{$pattern_{fine}$}

      $pattern_{fine} \gets \emptyset$ \;  
      \ForEach{$p_{i} \in pattern$}{
        $p_{children_{i}} \gets$ get all children nodes according $G$ \;
        \ForEach{$child \in p_{children_{i}}$}{
            
            originCost $\gets Cost(p_{i}, segment)$ \;
            $p_{fine_{i}} \gets p_{i}$ replace by child \;
            specificCost $\gets Cost(p_{fine_{i}}, segment)$ \; 
            \If{specifCost $>$ originCost}{
                $p_{i} \gets p_{fine_{i}} $ \; 
            }
        }
        $pattern_{fine} \gets pattern_{fine} \cup p_{i}$  \; 
      }
      \Return{$pattern_{fine}$}

     \caption{greedyTravel(pattern, segment, G)} \label{alg:fine_grained_travel} 
\end{algorithm}

 \section{Optimal update}

\subsection{Data augmentation}
\begin{figure}[h]
    \centerline{\includegraphics[width=0.45\textwidth]{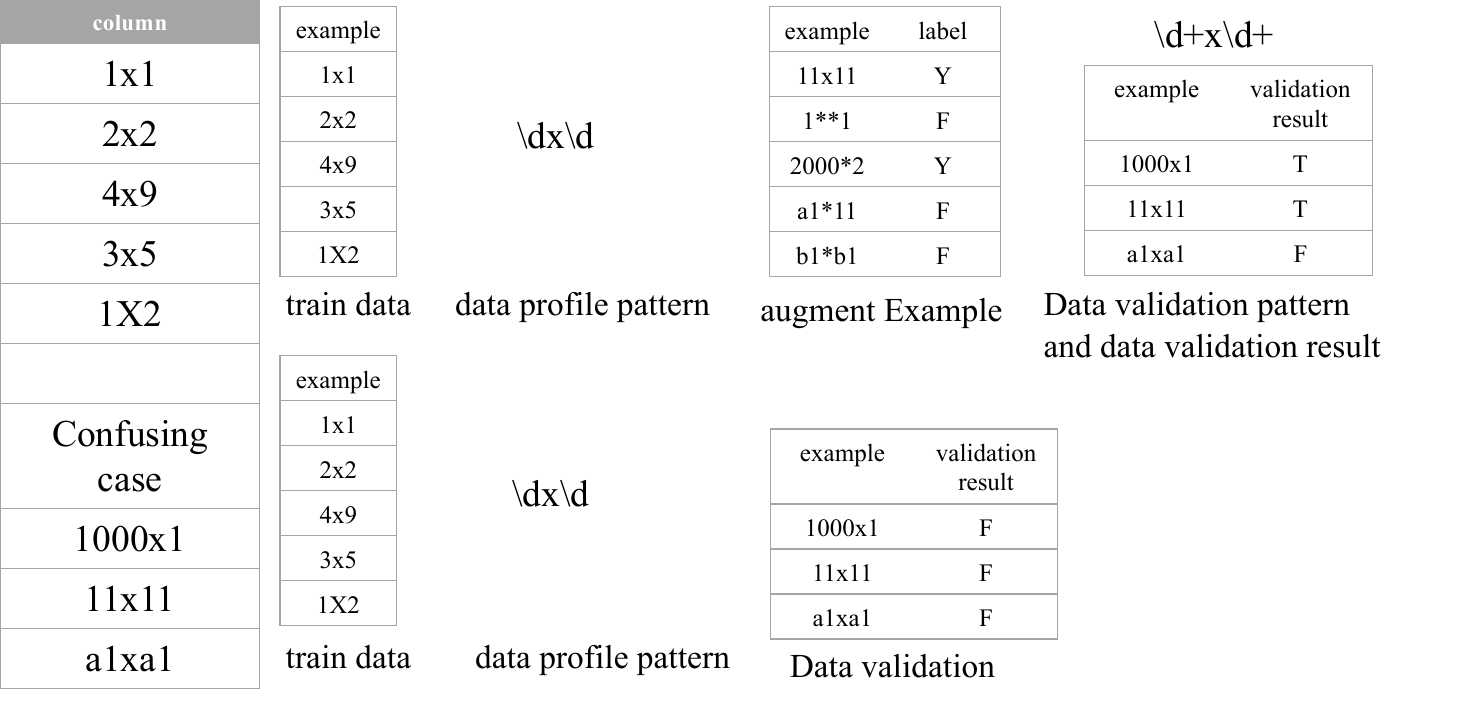}}
    \caption{example generation}
    \label{Fig.data augmentation}
  \end{figure}

{\color{black}
In reality, the best data profiling methods are often insufficient for direct data validation due to the limited historical data available, which typically captures only a small portion of the underlying data patterns. Our research has revealed that during the initial data insertion stage, data without sufficient features are easily identified as false-negative. Some methods, as outlined in \cite{song2021auto}, utilize corpora to compensate for the limitations of relying solely on historical data. However, the effectiveness of these methods is significantly influenced by the quality of the corpus, which is often difficult to obtain in practice. We therefore propose that human feedback during the pattern learning process can be utilized to enhance the accuracy of the patterns learned.

Figure \ref{Fig.data augmentation} illustrates the original training data, denoted by ``\verb|train data|'', which generates the pattern \verb|\dx\d|''. However, as depicted in the ``\verb|confusing case|'' data, determining the correctness of some data is often challenging. Subsequently, the data validation system intercepts such data during the verification process, leading to false-negative identification upon user confirmation. Frequent requests for user intervention can adversely impact the user experience. Thus, it is imperative to minimize the intervention cost at the initial stage. Therefore, we propose to introduce some methods to reduce subsequent intervention.
One intuitive approach to address the issue of cold start in data validation is to randomly generate data and prompt users to confirm their correctness. However, this method has limited effectiveness, as randomly generated data seldom represent boundary situations in pattern generation, and users are unlikely to recognize them as positive examples. This results in additional time required to seek user confirmation, without any significant improvement in algorithm performance. To overcome this challenge, we propose a method to automatically generate example data during pattern generation.

To enable automatic generation of example data during pattern generation, we propose to modify Algorithm \ref{alg:fine_grained_travel}. Specifically, we introduce a parameter $k$ to control the number of generated examples. After executing line 9, the algorithm checks whether the number of generated samples meets the input threshold parameter $k$. If not, it enters the example generation stage. The algorithm randomly selects nodes at the same level as the current refinement pattern, with the same parent node, as potential boundaries and replaces the original pattern using \textit{sample}($p_{children_{i}}\backslash child$). The algorithm generates data samples using enumeration and combination techniques. The process of transforming patterns into data is the inverse of transforming data into patterns. Characters can be mapped upwards on the generalization tree to enable character generalization. The atom pattern can be converted from the pattern to a specific character by searching down the generalization tree until it reaches the leaf node.}

\subsection{Incremental update}

\begin{figure}[h]
  \centerline{\includegraphics[width=0.35\textwidth]{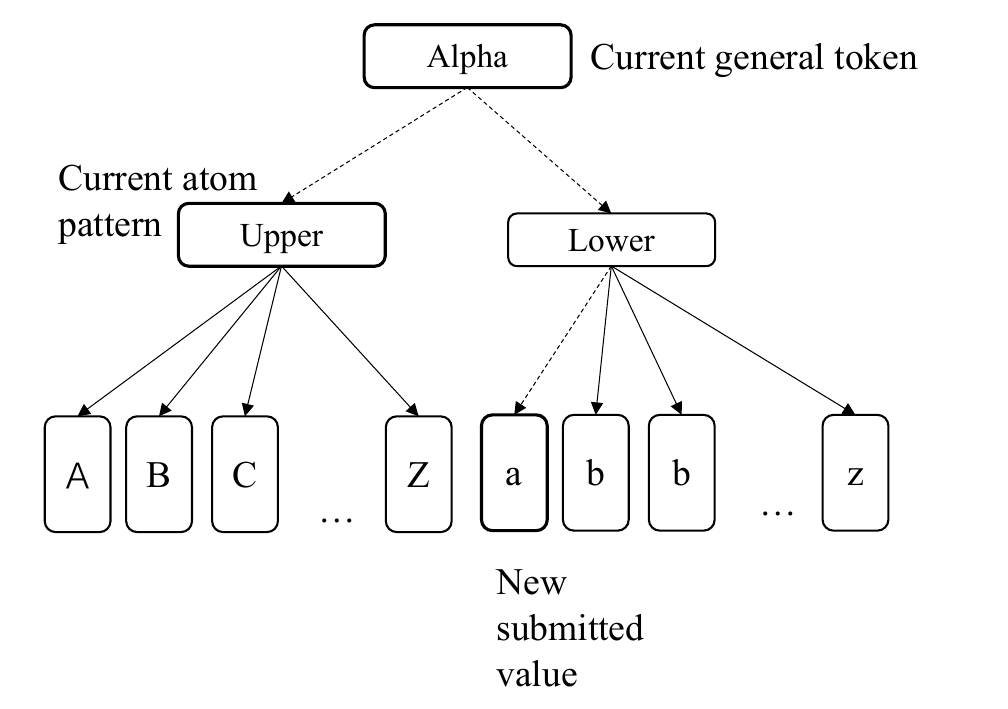}}
  \caption{Incrementally update pattern with submitted value}
  \label{Fig.increment}
\end{figure}

{\color{black}
\label{sec:inc}
In real-world scenarios, data is seldom static, and systems typically generate data continuously, or users submit data daily. As such, self-validation data management systems should be able to update constraints incrementally without accessing previous data. To this end, we propose a method for making incremental adjustments to patterns.

\begin{equation}
    \begin{split}
    P^{(t)} & = f(\Delta D{(t)}, P^{(t-1)}) \\
    \end{split}
\end{equation}

According to the above method, we specify the update stategy as follows.
There are two scenarios to consider when validating submitted data. If the submitted value is compatible with the constraints, there is no need to modify the pattern \textit{P}. However, if the submitted values violate the data validation constraints, the system will deploy generalization based on the new submitted data and the current pattern. The principle of incremental updates is to search on the generalization tree so that it can cover subsequent data. As illustrated in Figure~\ref{Fig.increment}, the pattern update algorithm solves the nearest common ancestor node of the current token and the inserted data as the updated token.}

 \section{Experiments}
The experiments are carried out on a computer with an Intel(R) Xeon(R) Gold 6326 CPU @ 2.90GHz processor and 128GB RAM running on Ubuntu 20.04.

\subsection{Experiment setup}
\textbf{Datasets.}
We build benchmarks using real data to evaluate the performance of our data validation system. We divide the data into the following three groups according to the source and characteristics of the data. 

\begin{itemize}

\item  \textbf{Kaggle} (214 dataset). The data of Kaggle dataset comes from Kaggle~\cite{kaggle} website. It consists of a lot of csv files. We use the data in each column as all values corresponding to a key. The final Kaggle dataset consists of a lot of different domains including name, date, zip code and so on.
 
\item \textbf{Random} (1000 dataset). We use a column in the Kaggle data set as an atomic unit, and randomly generate a random data set by randomly selecting special symbols. This dataset can evaluate the performance of our method on complex nested data.

\item \textbf{Enterprise} (1000 dataset). The data of enterprise dataset comes from the industry platform, which contains about two million key-value pairs. We randomly sample 500 pairs that contains different domain or different atomic domain to form enterprise dataset. 
\end{itemize}

\noindent{}\textbf{Methods.}
We compare the following methods.
\begin{itemize}

\item     \textbf{AutoPattern} is our proposed approach. We implement three variants of AutoPattern including AutoPattern-r, AutoPattern-v, and AutoPattern-rvf. AutoPattern-r is the implementation of recursive splitting in skeleton extraction. AutoPattern-v is the implementation of skeleton extraction. AutoPattern-rvf is the implementation of skeleton extraction and fined-grained semantic extraction. 

\item     \textbf{Deequ}~\cite{schelter2018automating, 
    Deequ} is a library built on top of Apache Spark for defining "unit tests for data", which measure data quality in large datasets. Deequ provides a lot of described constraints from developers to write customized constraints, it also can provide some constraint suggestions based on the data itself. 

\item      \textbf{FIDEX}~\cite{wang2016fidex} is a filter used to filter desired data from spreadsheets. Given some positive and negative examples, FIDEX will generate the proper pattern to describe the features of data.

\item      \textbf{TensorFlow data validation (TFDV)}~\cite{breck2019data,TFDV} can compute descriptive statistics that provide a quick overview of the data in terms of the features that are present and the shapes of their value distributions. We install TFDV via python-pip 
    and integrity it into the Python program.

\item      \textbf{xSystem}~\cite{ilyas2018extracting} proposed a branch and 
    merges strategy to pattern profiling. Its
    implementation~\cite{xSystem_code} is used
    to generate patterns. 

\item      \textbf{FlashProfile}~\cite{padhi2018flashprofile} is a tool for learning syntactic profiles for a
    collection of strings. It uses a set of regex-like patterns to describe the syntactic variations in the strings. Its implementation~\cite{FlashProfile_code} is used to generate patterns.
\end{itemize}

\textbf{Metrics.}
Given a dataset $D$, we evaluate the precision and recall of data validation as follows. We split each key of the dataset $D_k$ into two partitions and use the first 10\% of values as the training data to learn data constraints. To simulate the process of validating incoming values, we add some incorrect values into the second part and use them as testing data. The precision is defined as the percentage that correct data passed over all correct testing data and the recall is defined as the percentage of the incorrect value rejected over the incorrect values added.

Formally, our algorithm will observe training data and infer constraints. The inferred constraints will be tested on two groups of testing data: (1) the remaining 10\% of data, (2) the other pieces of data. In terms of (1), these testing data have the same potential domain as the training data, so the inferred constraints should not detect that these data are out of validation. Otherwise, it should be treated as a false-positive. In terms of (2), if the constraints inferred by our algorithm are too general, it is hard to detect the
data quality problem. So we also test the constraints in the other pieces of data.

\begin{figure*}[t]
    \centering  \subfigure[Kaggel dataset average precision and average recall]{
    \label{Fig.kaggle_PR}
    \includegraphics[width=0.32\textwidth]{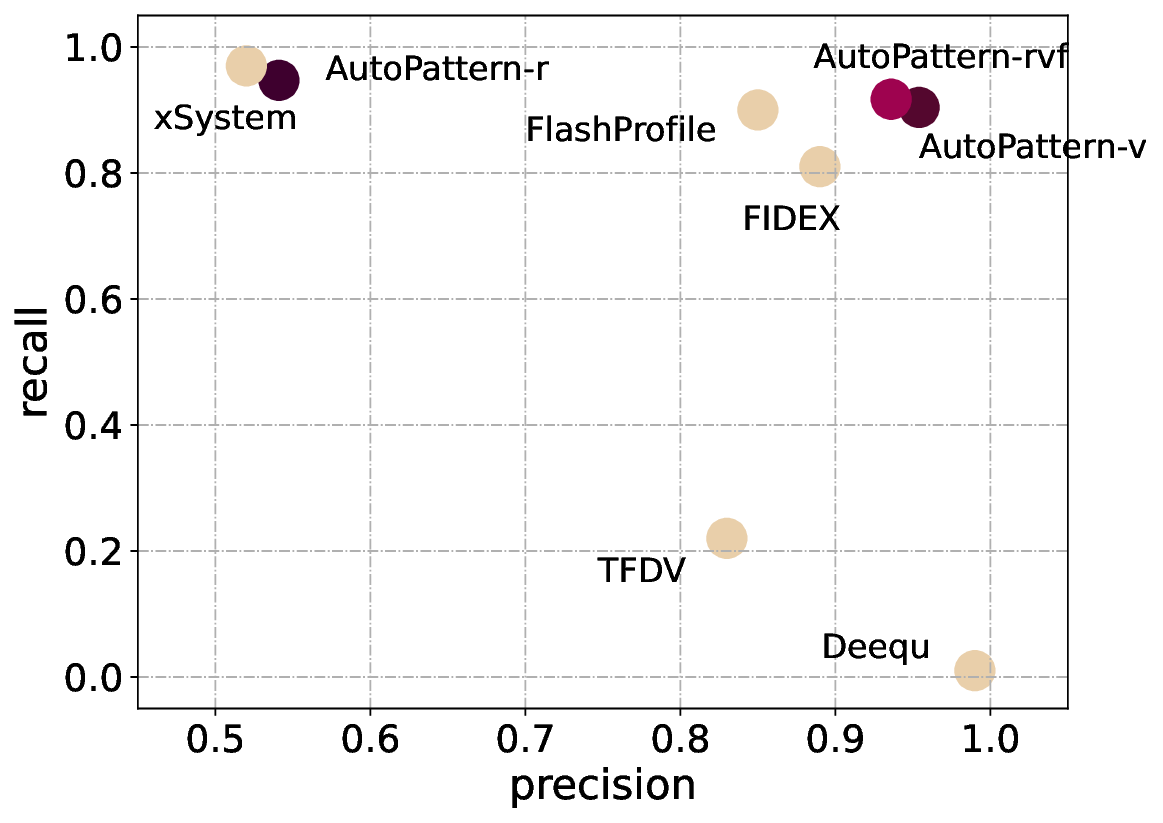}}
    \subfigure[Random dataset average precision and average recall]{
    \label{Fig.Random_PR}
    \includegraphics[width=0.32\textwidth]{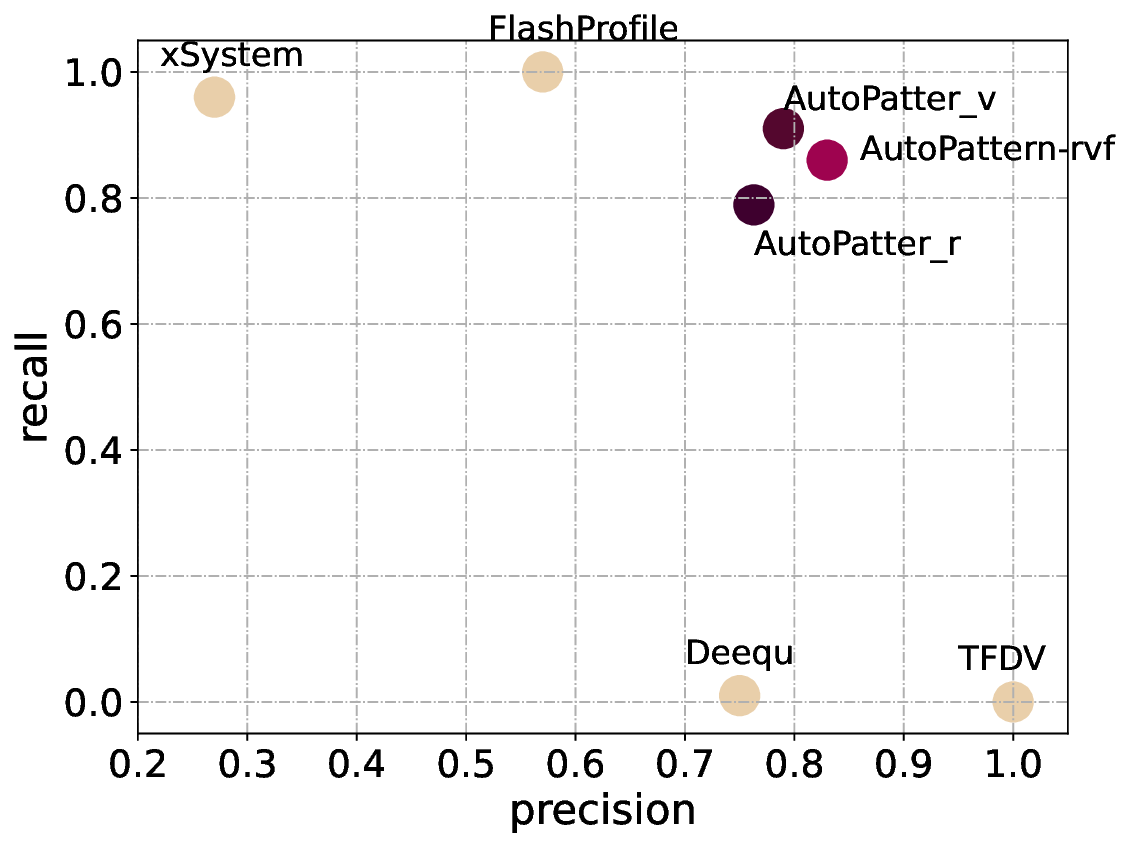}}
    \subfigure[Enterprise dataset average precision and average recall]{
    \label{Fig.Enterprise_PR}
    \includegraphics[width=0.32\textwidth]{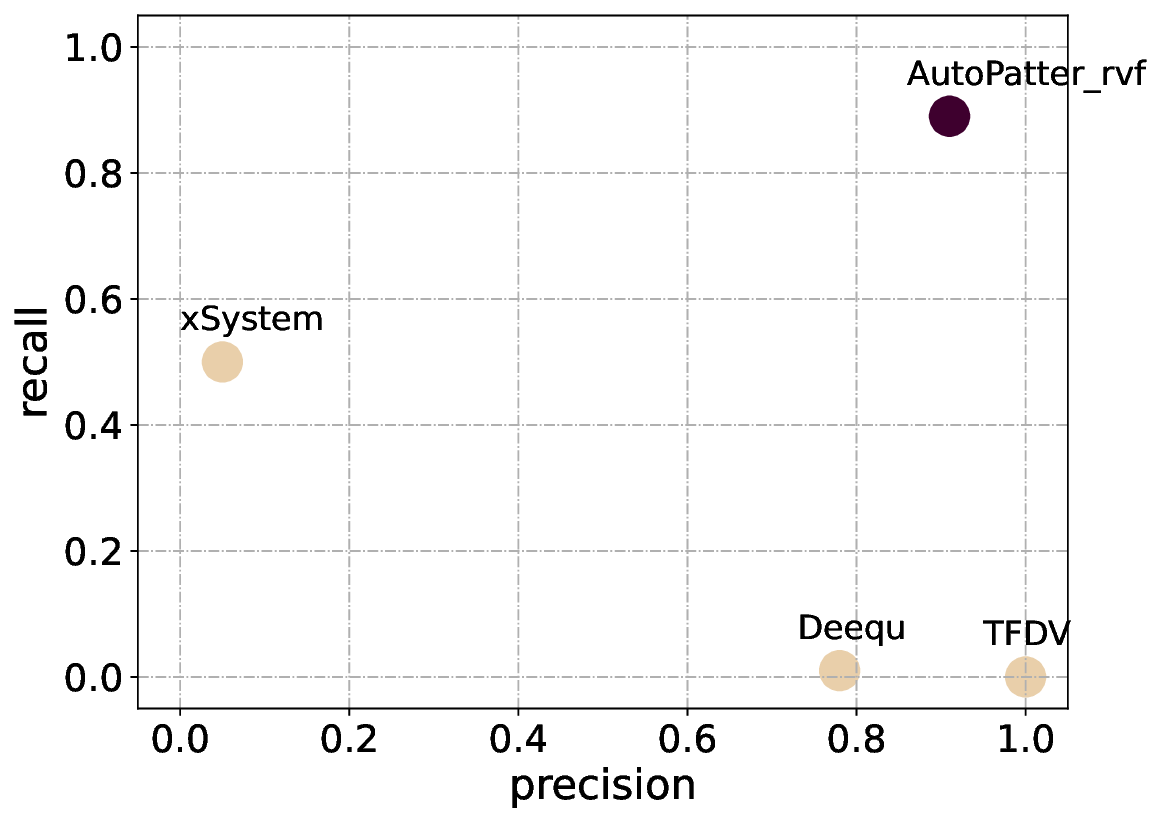}}
    \caption{Avg Precision and Recall.}
    \label{Fig:pr}
\end{figure*}

  \begin{figure*}[t]
  
    \subfigure[Sample rate precision.]{
    \label{Fig.sensitive.1}
    \includegraphics[width=0.15\textwidth]{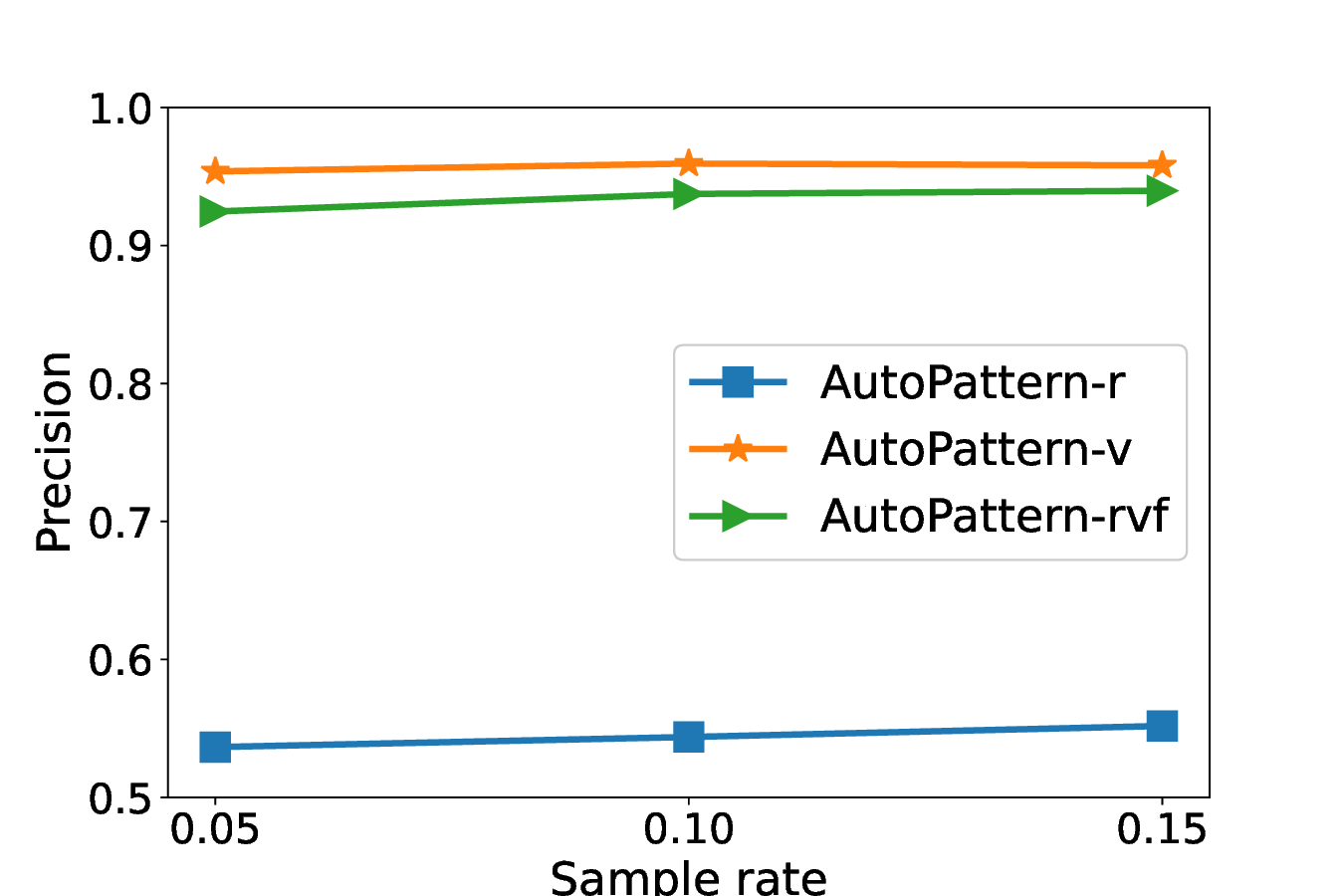}}
    \subfigure[Sample rate precision.]{
    \label{Fig.sensitive.1}
    \includegraphics[width=0.15\textwidth]{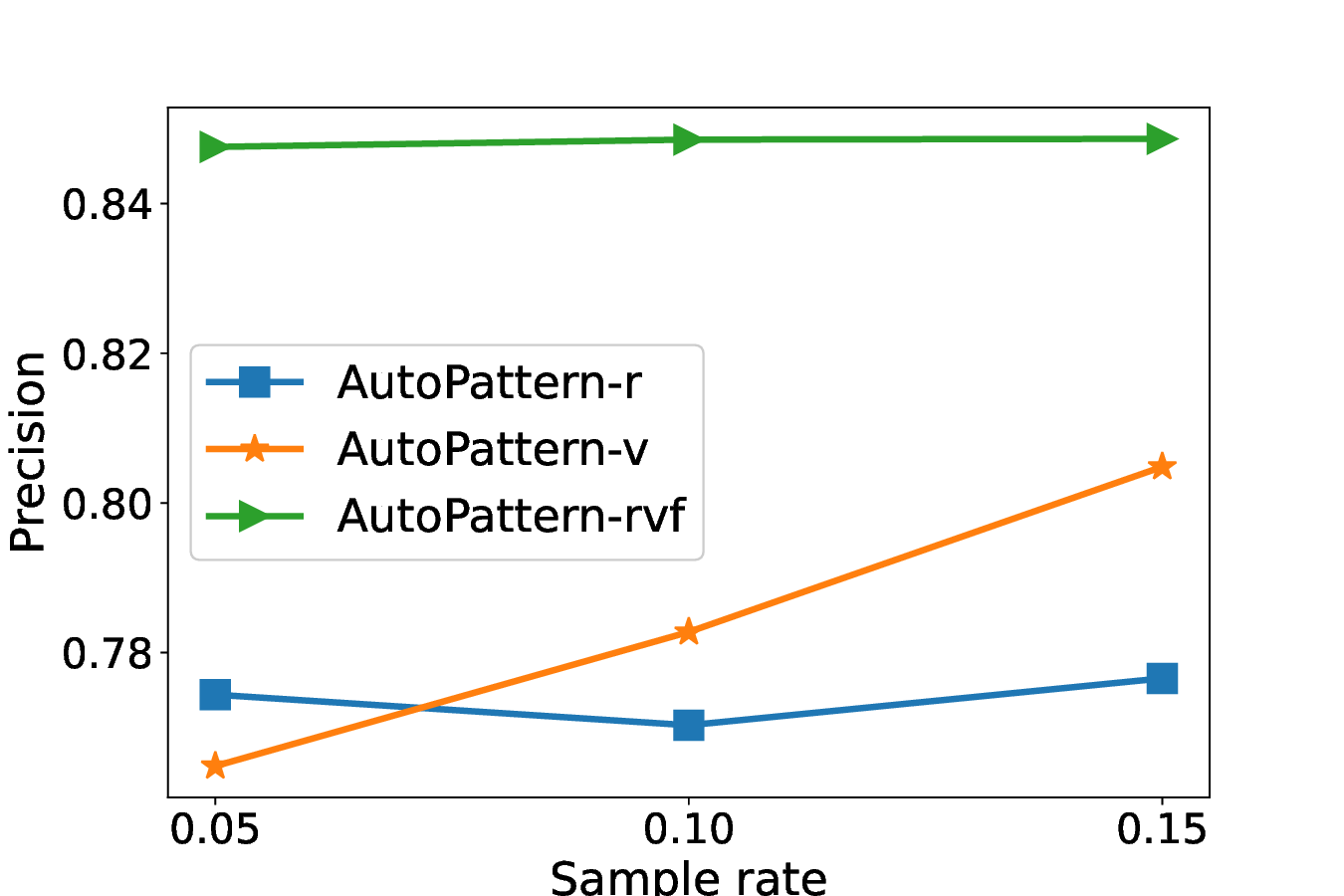}}
    \subfigure[Sample rate precision.]{
    \label{Fig.sensitive.1}
    \includegraphics[width=0.15\textwidth]{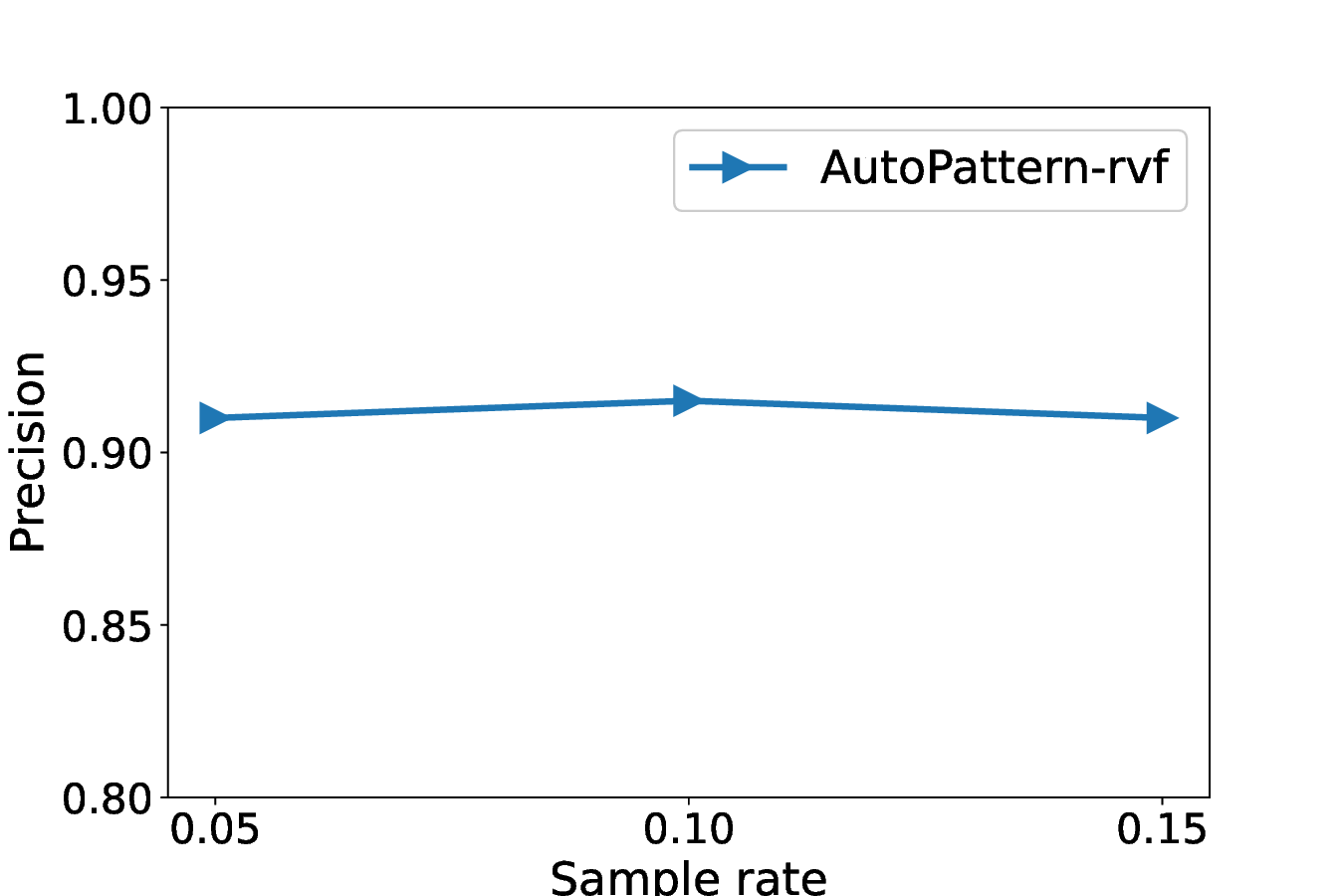}}
    \subfigure[Top-$k$ precision.]{
    \label{Fig.sensitive.1}
    \includegraphics[width=0.15\textwidth]{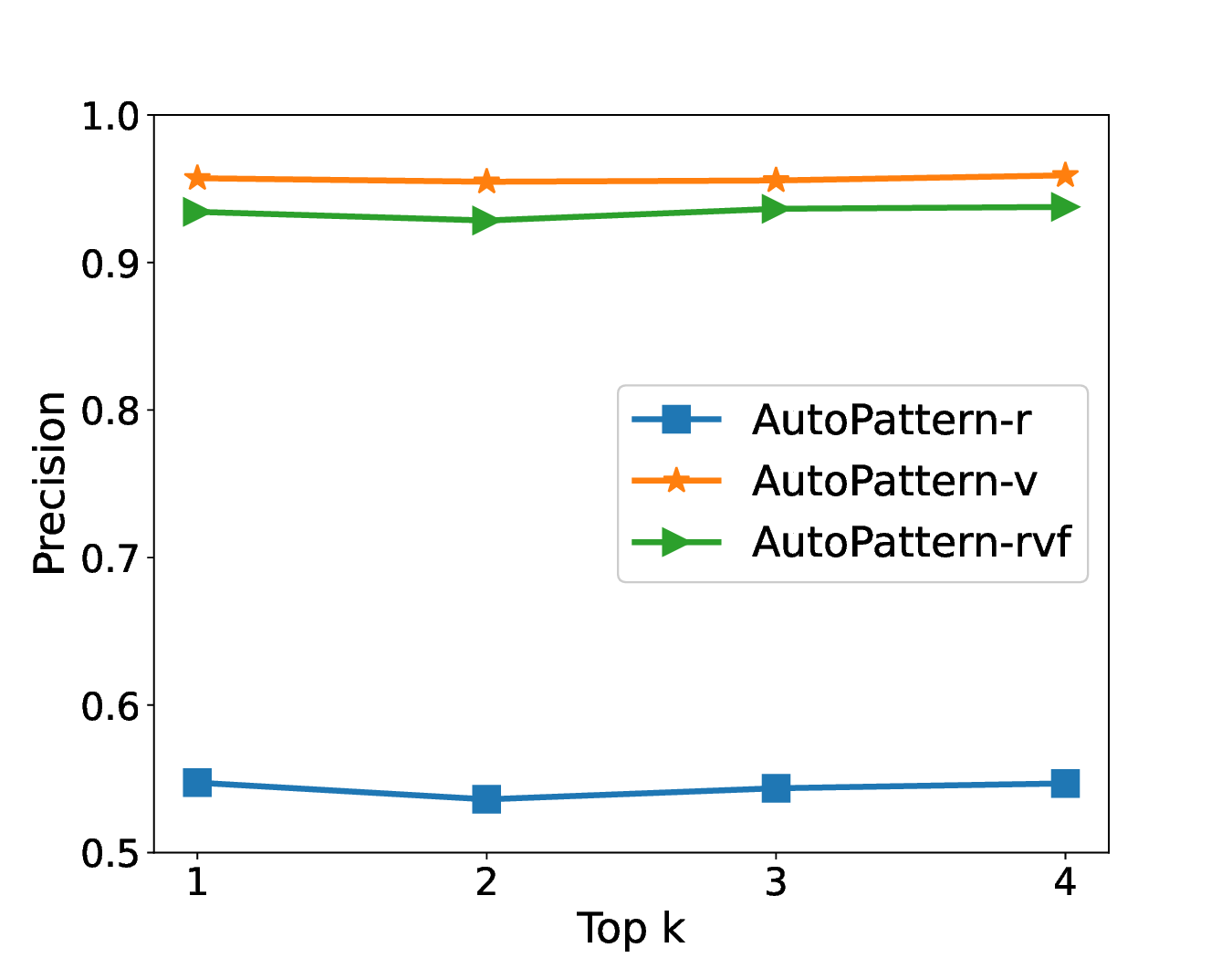}}
    \subfigure[Top-$k$ precision.]{
    \label{Fig.sensitive.1}
    \includegraphics[width=0.15\textwidth]{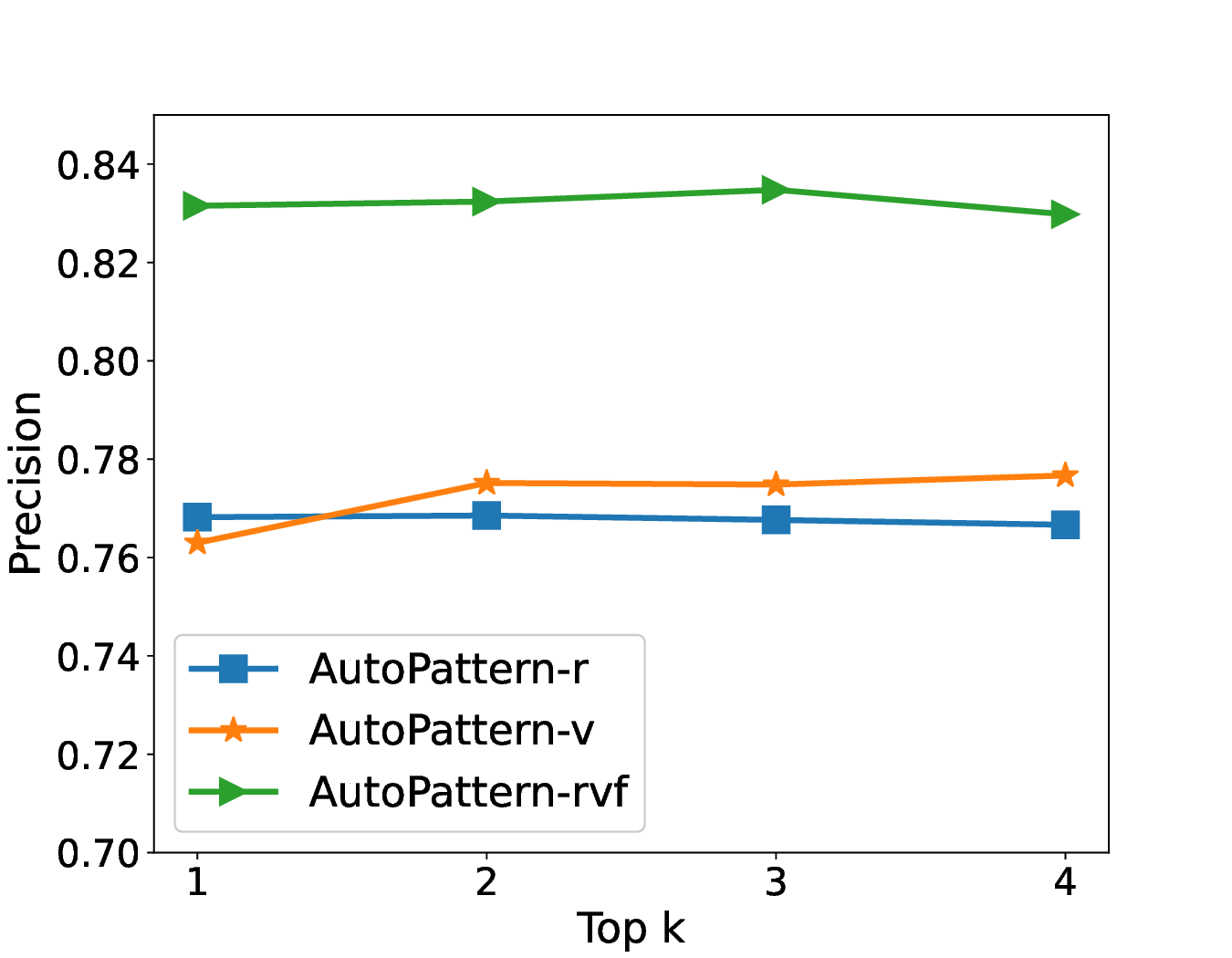}}
    \subfigure[Top-$k$ precision.]{
    \label{Fig.sensitive.1}
    \includegraphics[width=0.15\textwidth]{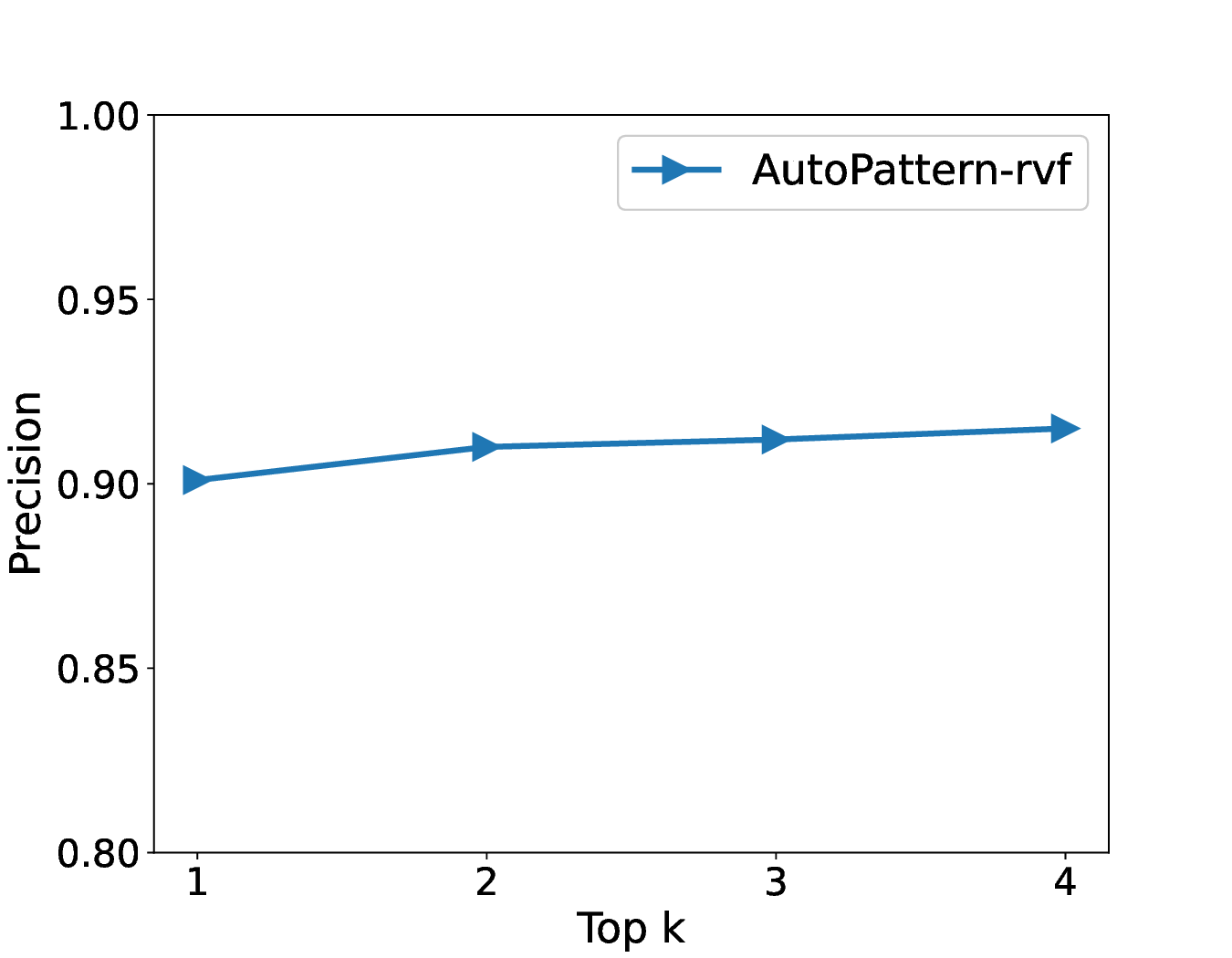}}
    \subfigure[Sample rate  recall.]{
    \label{Fig.sensitive.1}
    \includegraphics[width=0.15\textwidth]{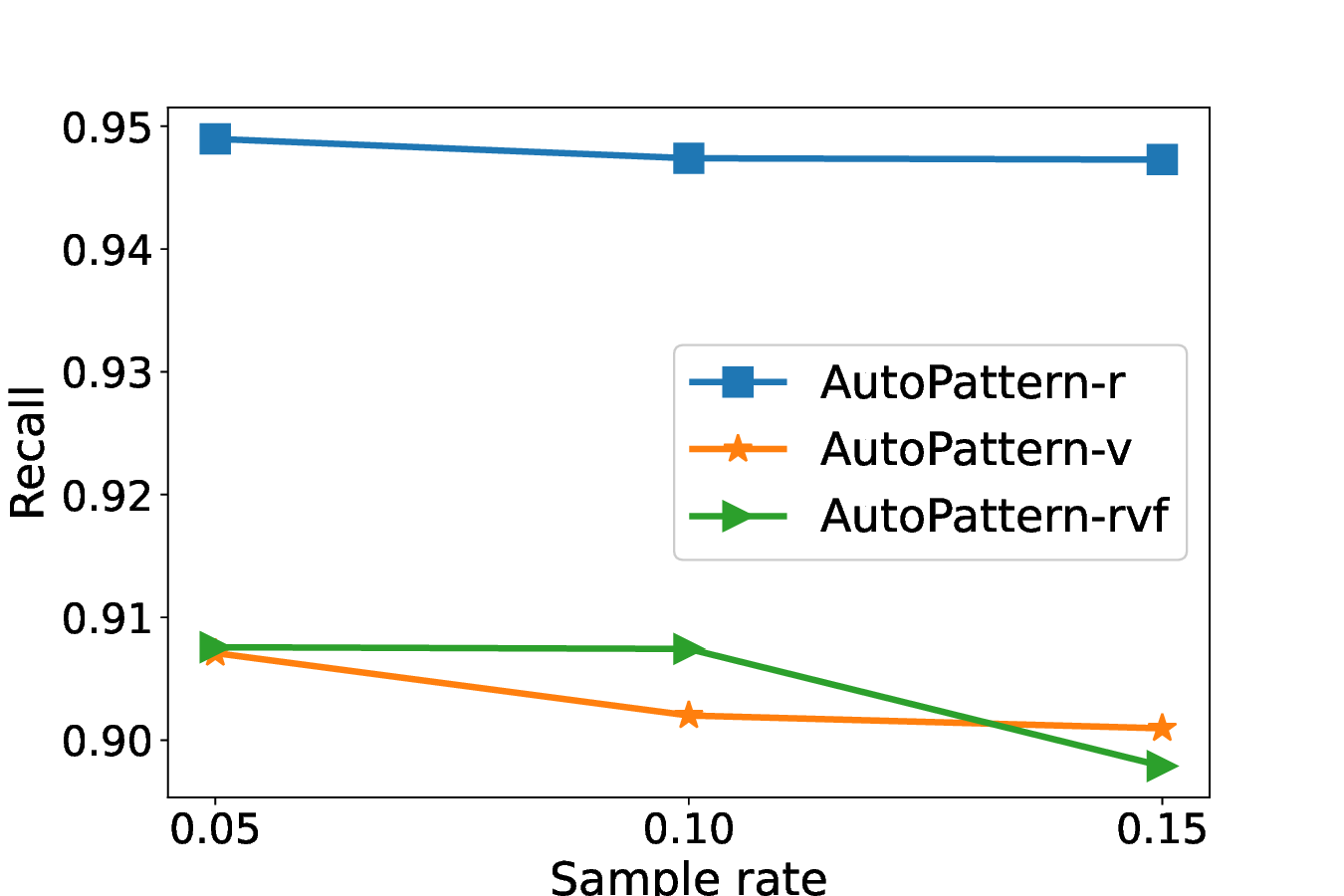}}
    \subfigure[Sample rate  recall.]{
    \label{Fig.sensitive.1}
    \includegraphics[width=0.15\textwidth]{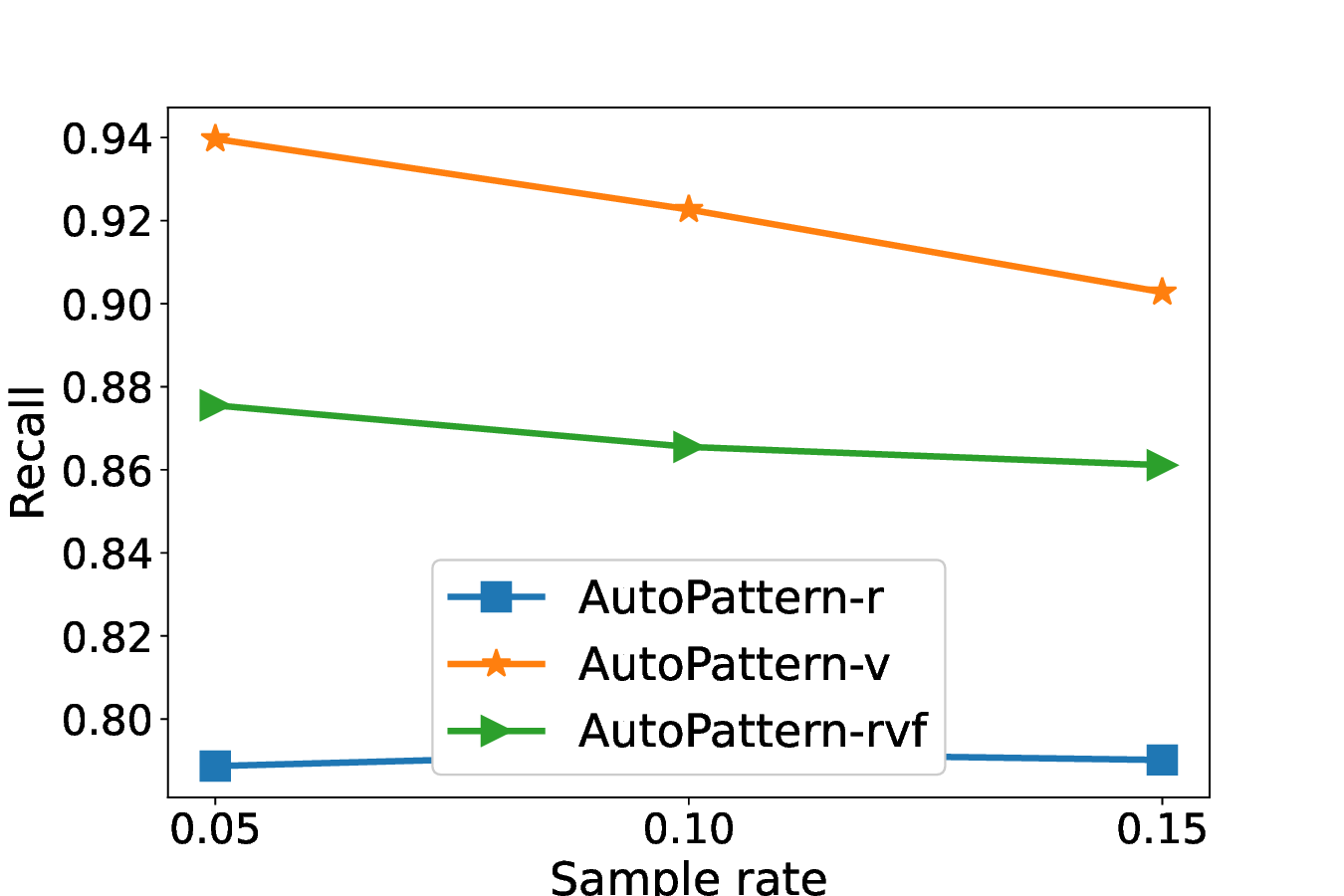}}
    \subfigure[Sample rate  recall.]{
    \label{Fig.sensitive.1}
    \includegraphics[width=0.15\textwidth]{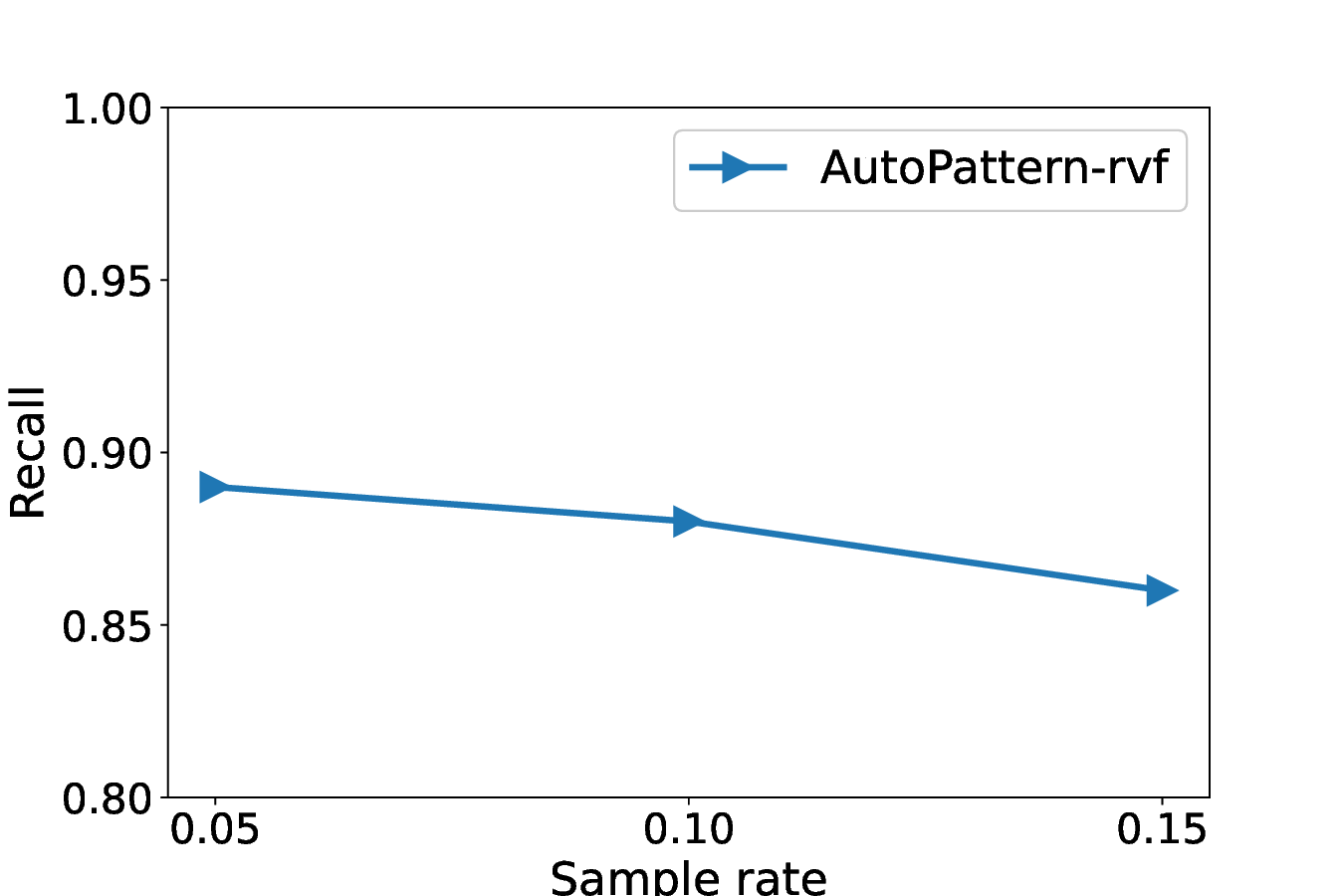}}
    \subfigure[Top-$k$ recall.]{
    \label{Fig.sensitive.1}
    \includegraphics[width=0.15\textwidth]{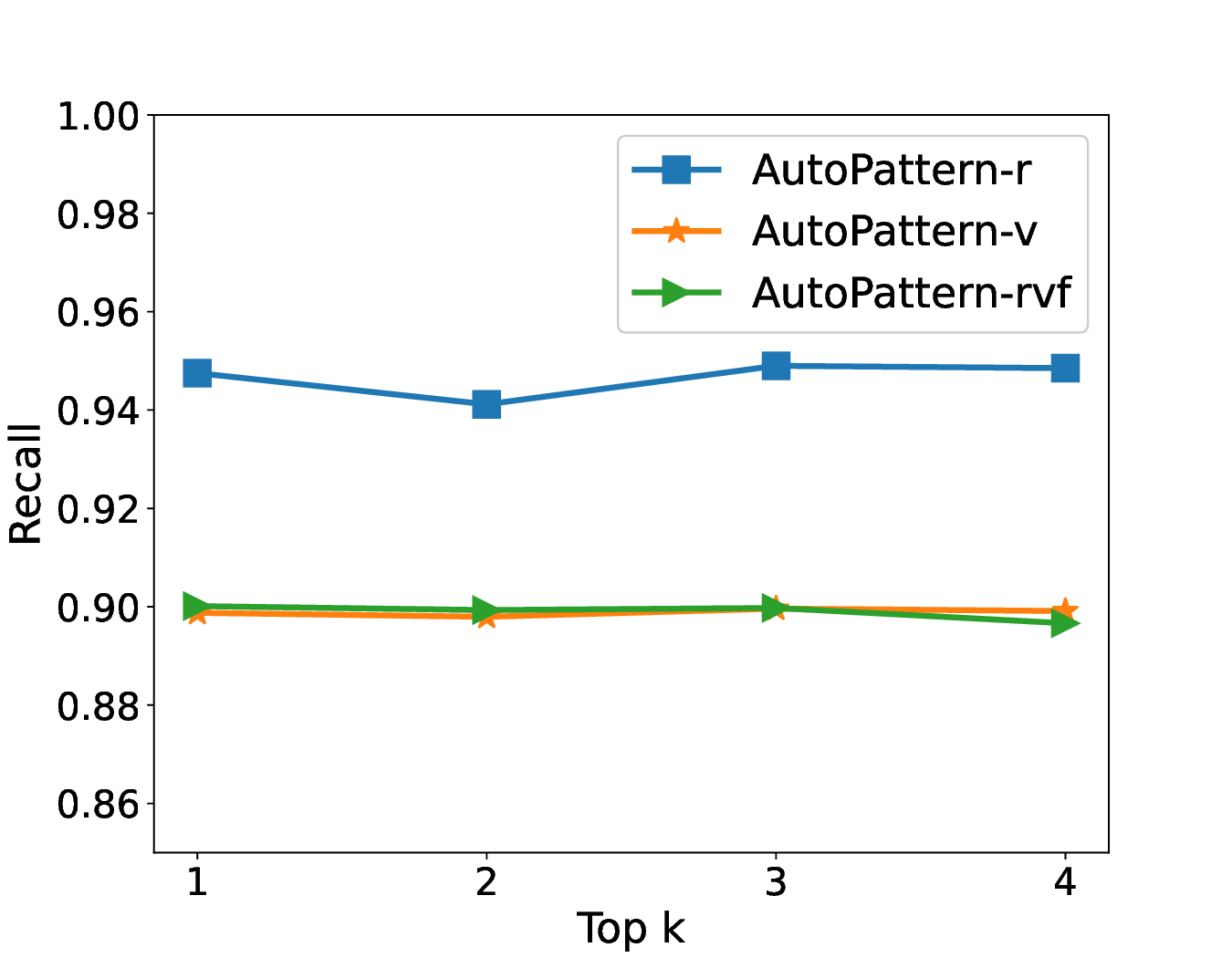}}
    \subfigure[Top-$k$ recall.]{
    \label{Fig.sensitive.1}
    \includegraphics[width=0.15\textwidth]{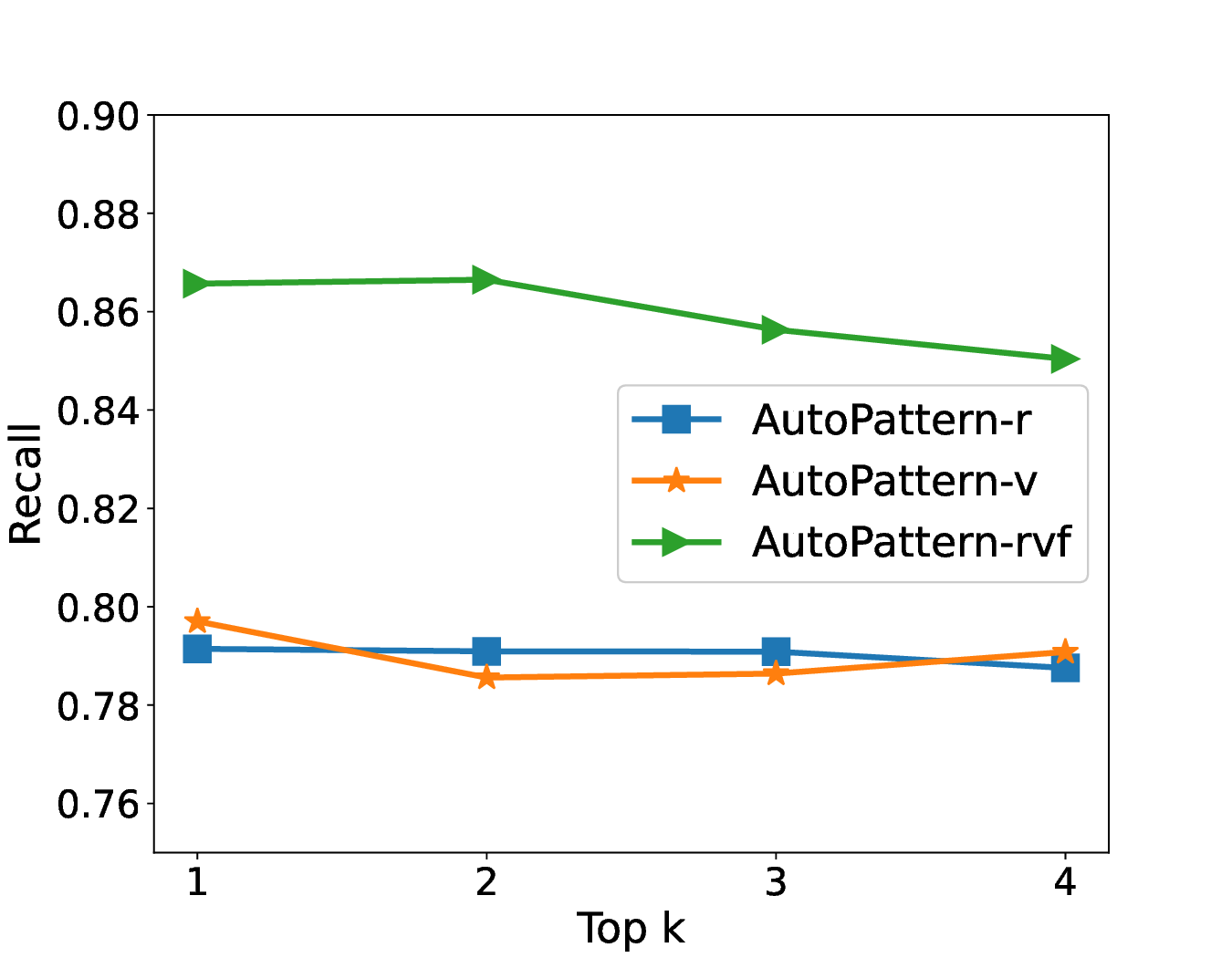}}
    \subfigure[Top-$k$ recall.]{
    \label{Fig.sensitive.1}
    \includegraphics[width=0.15\textwidth]{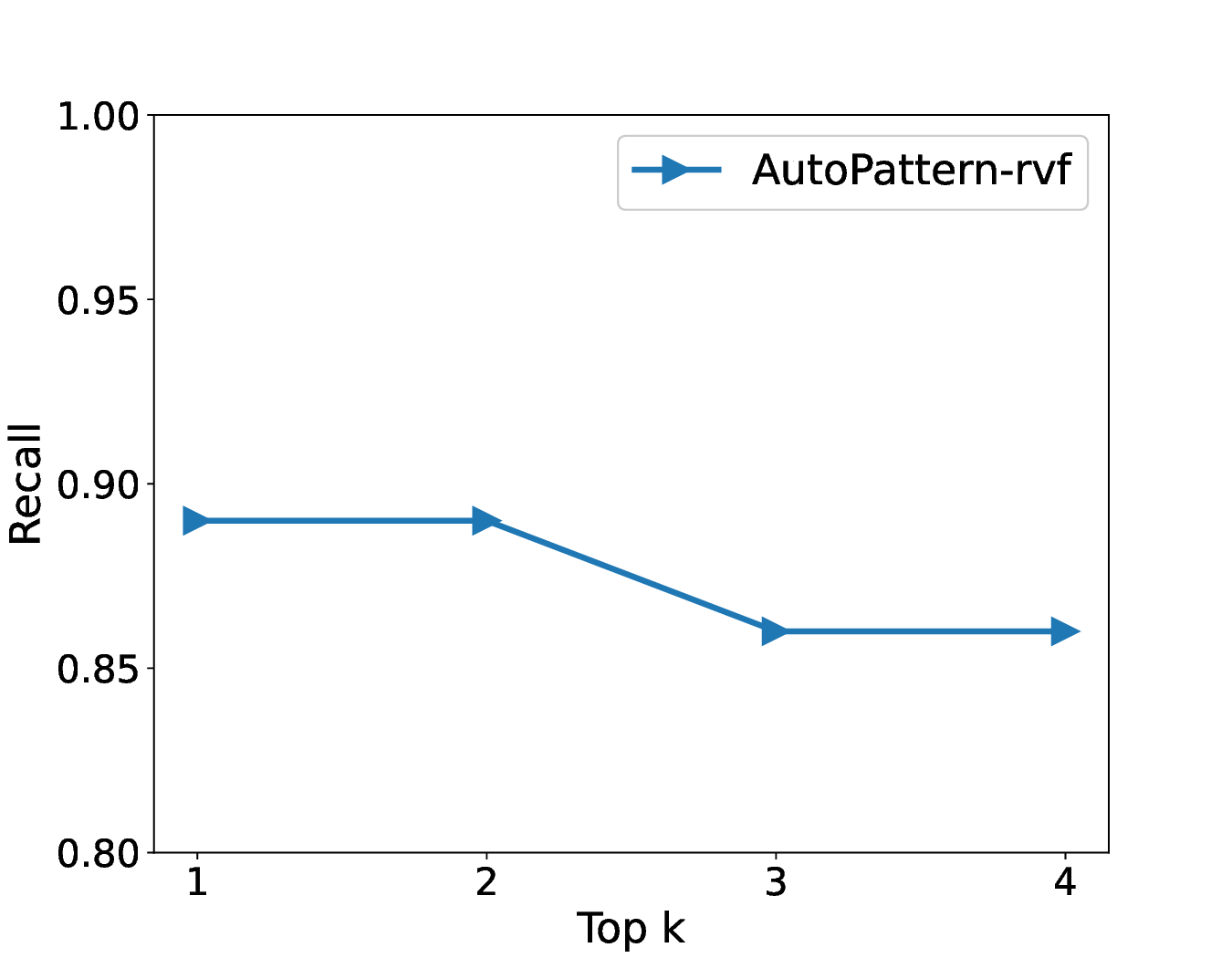}}
    \caption{Sensitivity}
    \label{Fig:Sensitivity}
  \end{figure*}

We now present the experimental evaluation focusing on the following key question. All methods are evaluated on the Kaggle and Random datasets. Limited by the industrial experimental environment in which the real-time data collected probably leads to performance loss, three methods (System, Deequ and TFDV) are tested on the Enterprise dataset. To our knowledge, it is sufficient to evaluate our proposed method in industrial scenarios.

\subsection{Precision and Recall} 

Fig.~\ref{Fig:pr} show the average precision and recall of AutoPattern compared methods in the three groups.
The experimental results show that the AutoPattern method is superior to the existing methods in the three datasets, which shows the effectiveness of AutoPattern in pattern generation.
Different AutoPattern variants perform differently in different datasets. In the Kaggle dataset, the accuracy of the AutoPattern-rvf method is lower than that of the AutoPattern-v method, but the recall rate is higher than that of the AutoPattern-v method, which shows that the fine-grained algorithm captures character-level semantics We realize that whether to perform pattern refinement involves the user’s trade-off between accuracy and recall. In the case of higher data accuracy and users can tolerate a certain degree of false interception, the system can be set to generate more refined pattern. In the Random dataset, the accuracy of the AutoPattern-rvf method has improved compared to the AutoPattern-r and AutoPattern-v variants. We realized that this is related to the nested structure of the Random data, using only recursive splitting or only using the vertical splitting is difficult to accurately capture the skeleton of the data in some scenarios, which leads to the failure of pattern generation, which makes the accuracy of the AutoPattern-rvf variant the highest among all comparison methods, which is different from the Kaggle dataset, because Kaggle datasets usually do not need to involve skeleton discovery and segmentation.

FlashProfile achieves the good performance in Kaggle dataset, but it performs not do well in Random
dataset, because FlashProfile lacks the inference of internal structure on complex nested
data, which generates patterns by copying the data in the dataset.  And FIDEX did not generate a pattern within the time limit in Enterprise dataset. 
That is because FIDEX use enumerate method to construct DAG (Directed acyclic graph) 
graph to represent the pattern of data. However, the data stored in the platform is usually very long 
(the average length is 939).
It also indicates that the previous method can not solve the complex string-valued data well. 
And the xSystem also uses a pattern-based method. But it also can not deal with long string-valued data. 
So it does not perform very well in the three datasets. 
The performance of FIDEX, FlashProfile, xSystem on different datasets illustrates the necessity of skeleton extraction.

\textbf{Scalability.}
This section mainly adjusts the predefined parameters in AutoPattern, focuses on the changes in the performance of AutoPattern under the changes of the predefined parameters, and evaluates the sensitivity of the algorithm.
There are two predefined parameters that need to be modified: (1) Sample rate (Sample rate): the amount of training data may affect the accuracy of the algorithm.
(2) Number of data patterns (Top-k): that is, the number of patterns used for data verification. The more patterns used for data, the more comprehensive the coverage of the pattern for historical data, but too many patterns can also have an impact on recall. This part of the experiment mainly observes the changes in precision and recall of the above two predefined parameters. We found that as the number of top-$k$ increases, the precision will increase and the recall will decrease.
In fact, the choice of the number of patterns used for data validation is related to the trade-off between recall and precision. But generally speaking, the change of parameters has no obvious influence on precision and recall.

\begin{figure}[t]
    \centerline{\includegraphics[scale=0.26]{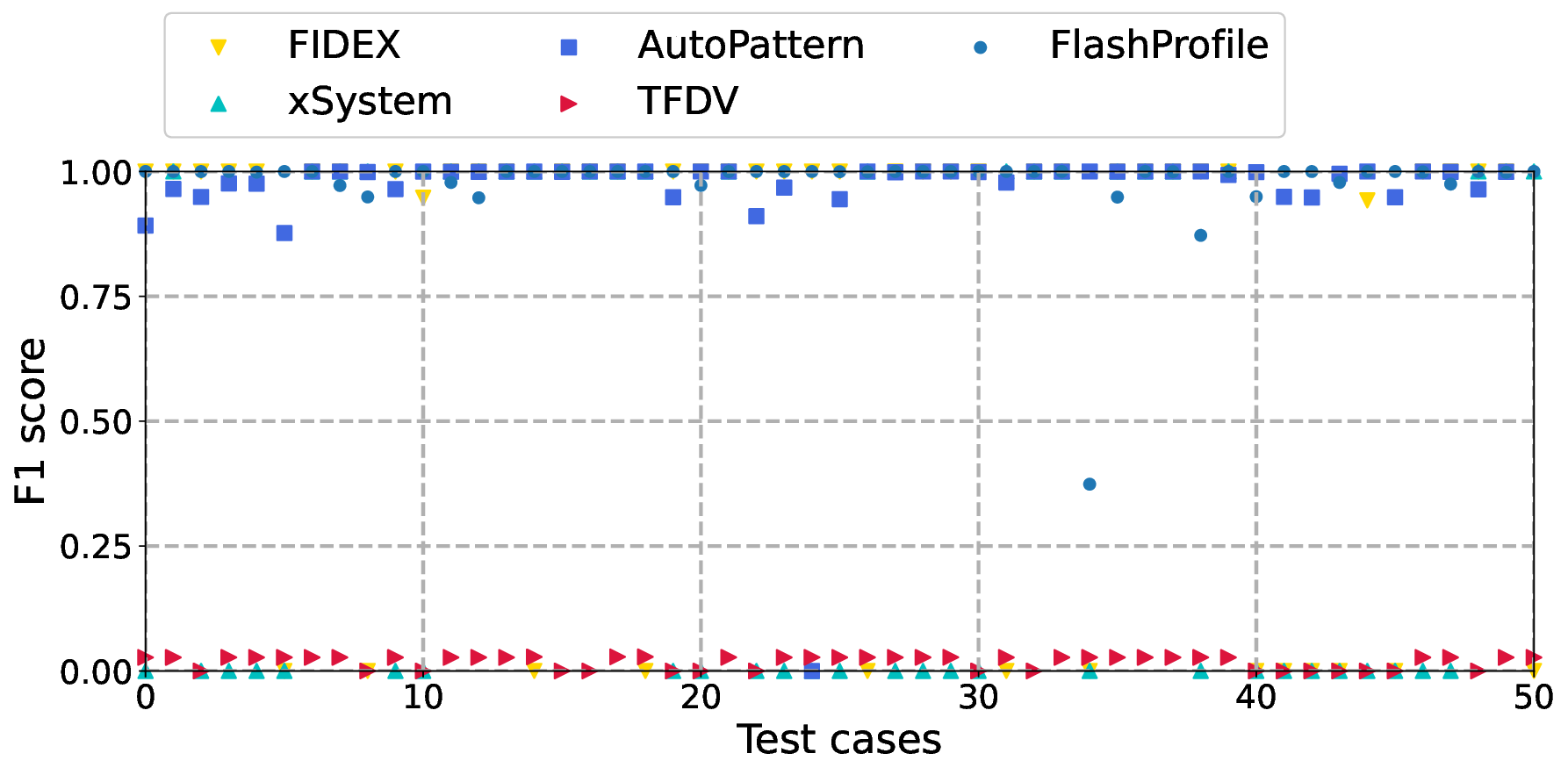}}
    \caption{Case by case results in sample data from Kaggle dataset.}
    \label{Fig.case by case}
  \end{figure}
  \begin{figure}[t]
    \centering  \includegraphics[width=0.5\textwidth]{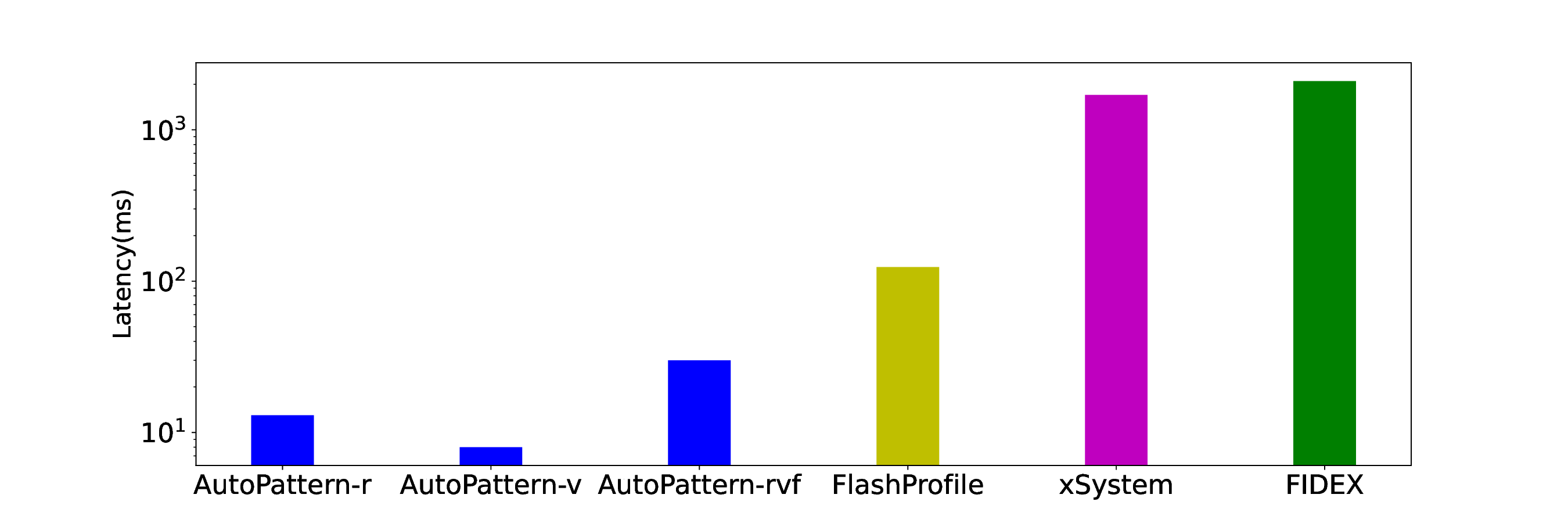}
    \caption{Average latency}
    \label{Fig.time latency}
  \end{figure}
  
\textbf{Case by case.}
We sample 50 cases and compute the f1 score to show the performance of different methods on a single case. The result in Fig.~\ref{Fig.case by case} shows that the pattern-based methods achieve better performance and our method achieves good performance in sample cases.

\subsection{User experience evaluation}

{\color{black}
The average latency of dataset processing to generate patterns is depicted in Figure ~\ref{Fig.time latency}. In the context of user interaction, low latency is crucial as excessive delay may lead to poor user experience. The proposed method takes a few seconds on average to process each dataset. Therefore, with low latency, even in multiple rounds of user interaction, the waiting time for the user in the data verification system will not be too high.
To assess the disparity in performance between automatically generated patterns and manually written patterns, we conducted an evaluation of the costs associated with manually writing data constraints. Specifically, we enlisted the participation of three programmers, each of whom possessed more than three years of programming experience and specialized in data science and engineering. Given their expertise in data conversion, filtering, and preprocessing using tools such as pandas, they spent a considerable amount of time in the preliminary learning stage, lasting several hours. While the learning speed may vary among individuals with different backgrounds, it is undeniable that automatic pattern generation algorithms provide a significant advantage for novices lacking a comprehensive knowledge base. With sufficient practice, our programmers were able to complete the majority of data constraint writing tasks within a few minutes.
The process of writing data constraints typically involves two key steps: (1) extracting a portion of the data preview and (2) crafting data constraints based on the extracted preview data while also ensuring their accuracy through multiple iterations. Consequently, the ability to automatically generate data constraints within seconds represents a substantial improvement over the time-consuming and labor-intensive process of manual constraint writing.}

\subsection{Data augment evaluation}
{\color{black}
In this section, we evaluate the effectiveness of incorporating user feedback into the AutoPattern's automatic pattern generation algorithm to simulate the dynamic process of data insertion and validation. Our main objective is to investigate the amount of human intervention required during the entire data verification process.

To conduct our experiment, we randomly selected 10 datasets from the distribution of the unique ratio $<$20\%, 20-80\%, and $>$80\%. We then performed two experiments: AutoPattern-sort method and AutoPattern-random method. In AutoPattern-sort method, we sorted the data in the dataset, selected the first 10 data as the initial training data, and used the subsequent data to simulate the insertion process. In AutoPattern-random method, we randomly chose 10 data as initial training data. Table~\ref{fig:data augment} summarizes the experimental results. The Autopattern method denotes generating patterns using an unsupervised approach, while the data augmentation method refers to adding data augmentation samples from user feedback in the first three rounds of patterns generation.
\begin{table}[t]
  \caption{The superscripts used in the table represent different data distributions with unique Rate of 0-0.2, 0.2-0.8, and 0.8-1.0. The table shows that different data distributions and different AutoPattern variants achieve different given precision thresholds ( 0.8 and 0.9 respectively) the proportion that requires user intervention (intervene) and the proportion that requires the use of training data (train)}
  \centering
  \small
  \resizebox{0.47\textwidth}{!}{
  \begin{tabular}{lccccc}
      \toprule 
      & \multicolumn{2}{c}{threshold=0.8} & \multicolumn{2}{c}{threshold=0.9} \\
      \cmidrule(lr){2-3}  \cmidrule(lr){4-5} 
      method &  intervene(\%) &  train(\%) & intervene(\%) &  train(\%) \\
      \midrule
      AutoPattern-sort$^{0-0.2}$ & 3.6 & 25.3 & 3.9  & 28.7 \\
      AutoPattern-random$^{0-0.2}$ & 0.0 & 1.1 &	0.0	 & 1.5 \\
      dataAugment$^{0-0.2}$ & 3.0 & 21.0 &  3.0 & 22.4 \\ 
      AutoPattern-sort$^{0.2-0.8}$ & 8.6	& 33.3	& 8.6	 & 36.0 \\
      AutoPattern-random$^{0.2-0.8}$ & 18.3 & 7.0 &	25.1 &	7.0 \\
      dataAugment$^{0.2-0.8}$ & 6.0	& 13.0	& 7.2	& 15.5 \\ 
      AutoPattern-sort$^{0.8-1.0}$ & 27.3 &	14.0 & 29.5 & 15.7 \\
      AutoPattern-random$^{0.8-1.0}$ & 7.8 &	1.3 &	9.1 &	2.1 \\
      dataAugment$^{0.8-1.0}$ & 15.1 & 11.1 & 17.5  & 13.3 \\
      \bottomrule
  \end{tabular}}

  \label{fig:data augment}
  \end{table}
  The table presents the required level of human intervention for simulating the data insertion process across different data distributions, as well as the amount of training data needed for the generated model to achieve a certain threshold of accuracy. The results indicate that data augmentation significantly improves the performance of the AutoPattern method when the unique rate distribution of the data is between 0.2 and 0.8, as well as when it exceeds 0.8. Conversely, when the unique rate value is too small and the data is skewed towards enumeration types, the data augmentation algorithm may struggle to generate appropriate augmented samples, leading to less significant improvements in performance. Under other data distributions, the AutoPattern method with data augmentation significantly reduces the required level of user intervention and the amount of training data needed to reach the accuracy threshold. The data augmentation technique increases the feature space of the training data, mitigating the risk of data locality and reducing the number of manual interventions in the system.}

\subsection{Data quality impact assessment}

\begin{figure}[h]
  \centering  \subfigure[gro\_data dataset]{
  \label{Fig.DQ.1}
  \includegraphics[width=0.22\textwidth]{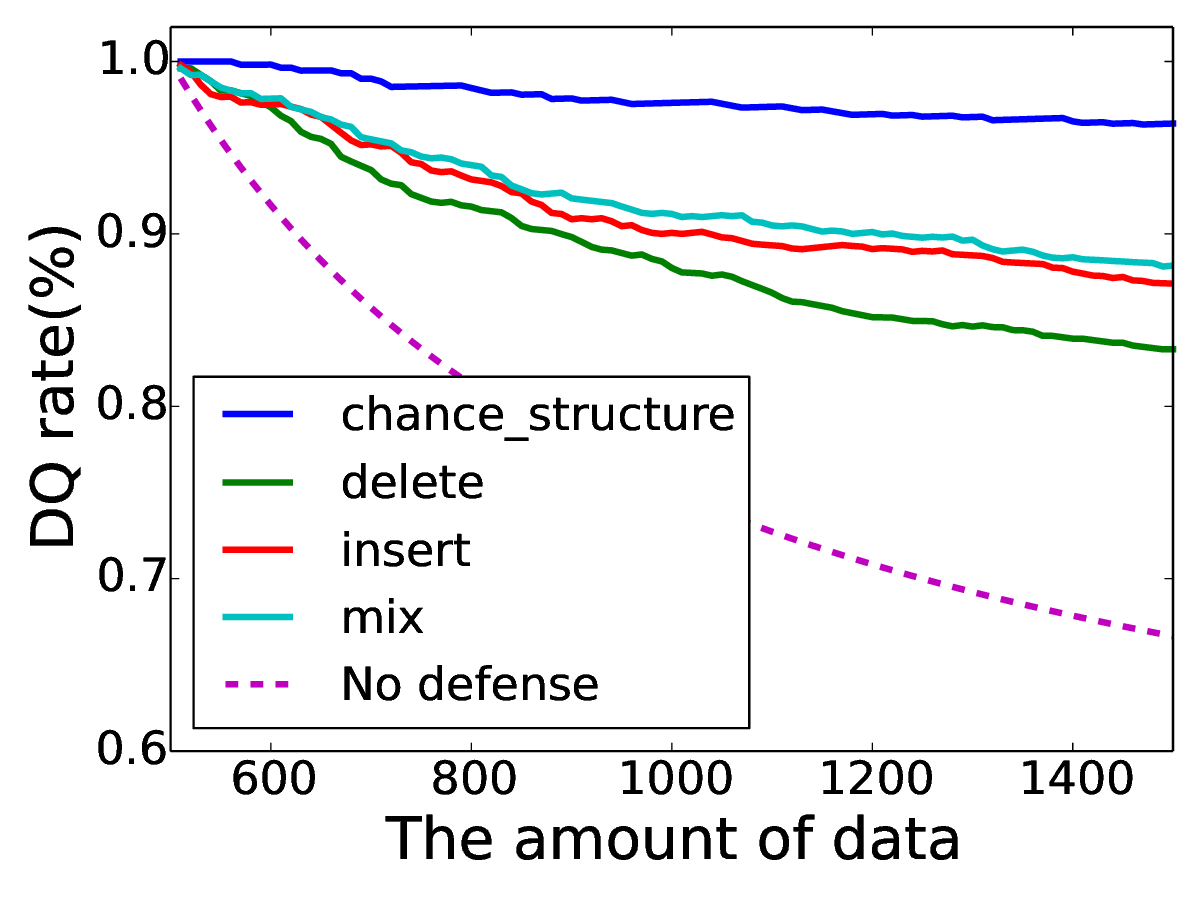}}
  \subfigure[market dataset]{
  \label{Fig.DQ.2}
  \includegraphics[width=0.22\textwidth]{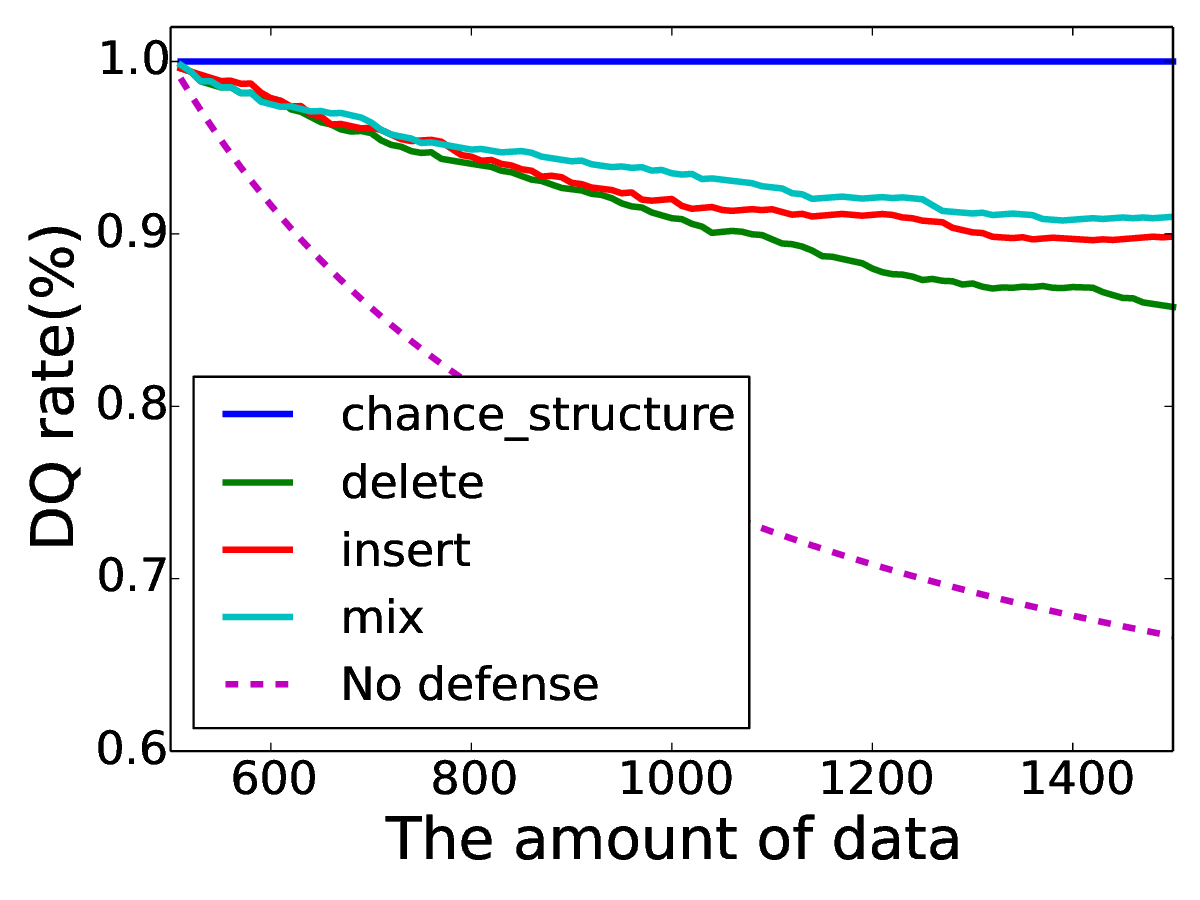}}
  \caption{The chance of DQ rate}
  \label{Fig.DQ}
\end{figure}
{\color{black}
We have designed a self-validate data management system to address the issue of declining data quality in unrestricted systems, which tend to accumulate increasingly large amounts of dirty data over time. In order to measure the effectiveness of our system, we assess data quality (DQ) by calculating the percentage of clean historical values among all the historical values. 

To quantify the validation performance, we perform the following simulation. 
We randomly sample 500 rows of data and manually mark them to ensure the correctness of the data. That is to say, we guarantee that the starting data set is clean.
Subsequently, we use 10 as batch size, five of which are clean data sampled from the original dataset
The other 5 are dirty data generated by modifying clean data. The modification method we use is as follows:
    (1) chance the structure of the data. We notice that if we change the delimiter, the structure of the data will be changed too; 
    (2) delete some characters; 
    (3) insert some character.

Without proper data validation and management, the influx of dirty data can have severe consequences on data quality. This is especially true in extreme cases where the data becomes unmanageable. To address this issue, we developed a data validation system using the AutoPattern method and evaluated its performance through simulations. The results, presented in Fig.\ref{Fig.DQ.1} and Fig.\ref{Fig.DQ.2}, indicate that the system is sensitive to structural changes in the data. Such changes are particularly risky, as they may cause downstream data processing programs to throw exceptions and interrupt the task flow. The impact of dirty data caused by changes at the character level is also shown, with the green line representing character insertions and the original line representing character deletions. The red line depicts the impact of mixing these error types on data quality. Notably, the dotted line illustrates the worst-case scenario where no data validation or management is implemented.}

\subsection{Case study}
{\color{black}
To demonstrate the real capabilities of the data validation system with 
AutoPattern, we integrate the data validation system into an industrial 
platform at Ant Group Inc. The system covers the configurations change scenarios in the industry. 
There are about two million configurations on the industrial platform.
These configurations are centrally stored in the database, and the database saves the history of all configurations.

We compare the pattern learned by our AutoPattern method with the outputs from 2 state-of-art tools: (1) FlashProfile~\cite{padhi2018flashprofile}, (2)Ataccama One~\cite{Ataccama_ONE}: a state of art data wrangling platform and their data validation results.
As shown in the Fig.~\ref{Fig.pattern profile.1}, FlashProfile captures the underlying patterns of the data very well, but unfortunately, this method treats the data as a whole and does not consider the internal nesting structure. Moreover, the data profiles method cannot predict the feature of the data that did not appear, so both positive and negative examples in the new data will be intercepted. Ataccama One finds the skeleton of the data and automatically divides one column of data into three columns. But unfortunately, for each column after division, Ataccama One method can only recognize some predefined types such as datetime. Both the first and third columns of the Fig.~\ref{Fig.pattern profile.1}(2) failed to learn the underlying patterns, 
only using enumerate patterns to cover all data. Moreover, Ataccama One focuses more on transforming the data without considering the structure of the original data. 
AutoPattern correctly captures the underlying skeleton structure, and the resulting pattern contains fine-grained semantics so that it can be used in data validation. 

As shown in the Fig.~\ref{Fig.pattern profile.2},
FlashProfile fail to capture the underlying pattern,
Ataccama One find the nested structure inside the data, but the performance of text segmentation is not well. AutoPattern correctly captures the underlying recursive skeleton structure. Therefore, our method can properly generate meaningful patterns, and the generated patterns can be used for data validation better than other methods.

\begin{figure}[h]
    \centering  \subfigure[Vertical skeleton data validation]{
    \label{Fig.pattern profile.1}
    \includegraphics[width=0.5\textwidth]{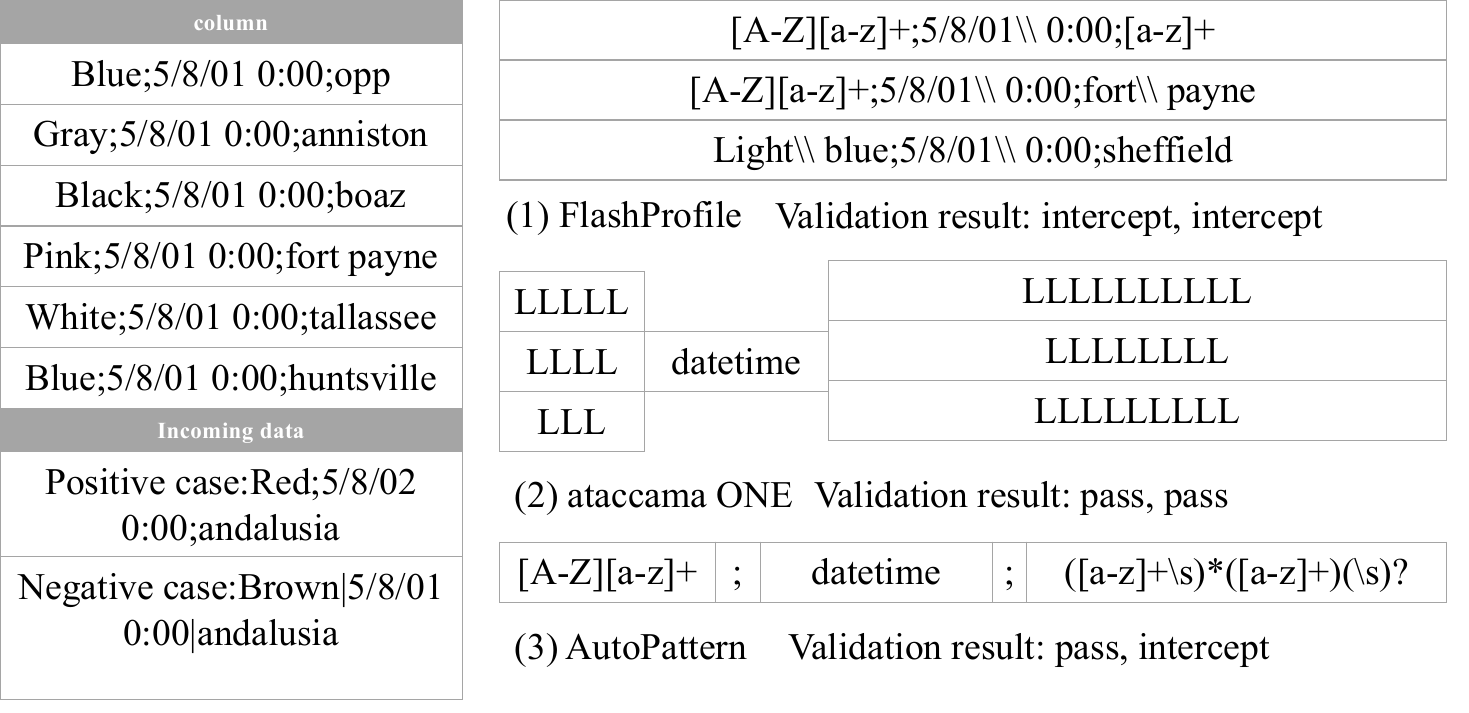}}
    \subfigure[Recursive skeleton data validation]{
    \label{Fig.pattern profile.2}
    \includegraphics[width=0.5\textwidth]{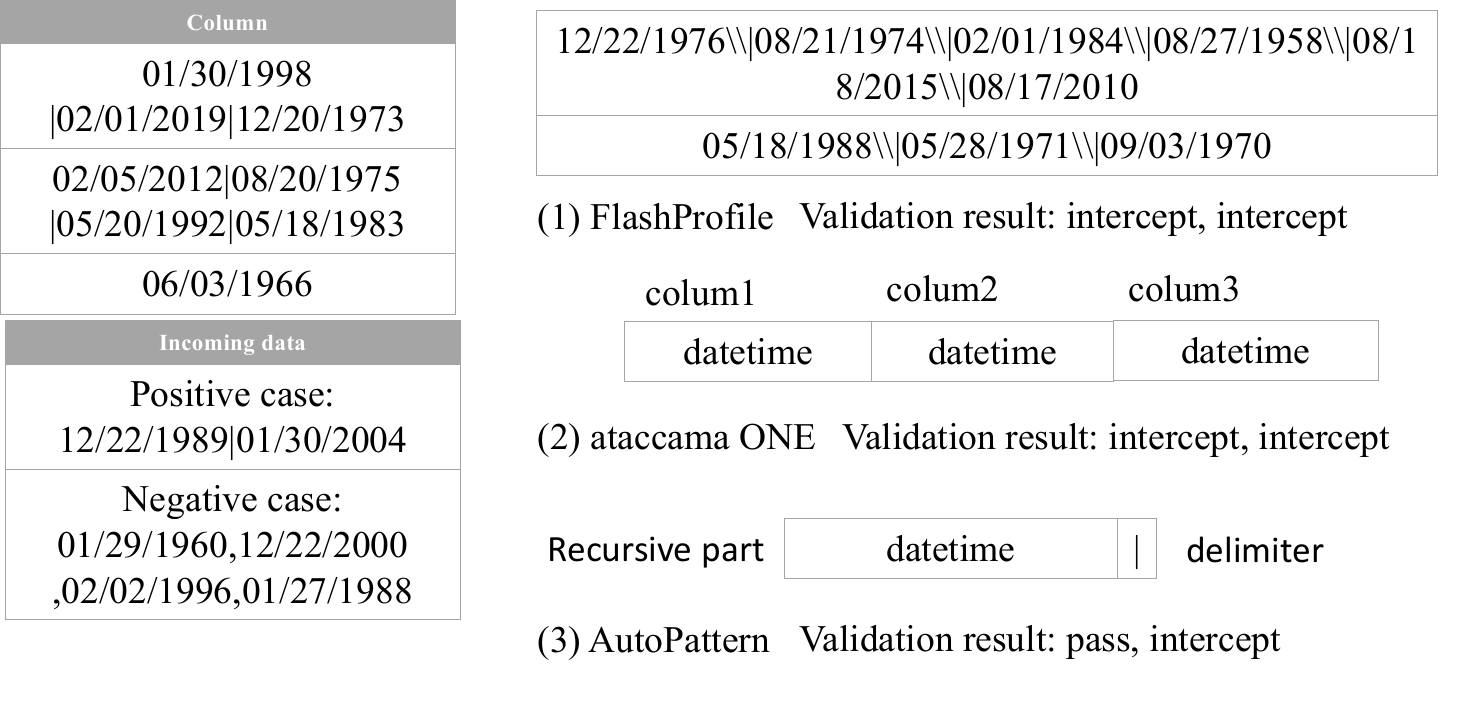}}
    \caption{Data validation case study, Validation results show the results of data validation using difference method in figure to validate incoming positive and negative cases. }
\end{figure}

The data validation has been deployed on the platform for several months 
and has been proven to produce effective interception for report error early and help engineer locate error effectively. Before deploying the data validation system, 
there have been many service crashes on the platform due to incorrect data. Moreover, data validation patterns 
can only be written manually based on human experience. And if the data 
validation pattern is manually written, the correctness of the data validation pattern also 
requires additional verification to ensure. Moreover, 
the data validation pattern needs to be rewritten again once the pattern 
drift occurs, which is quite time-consuming and laborious.}

 \section{Related work}

\textbf{Structure extraction.}
Structure extraction is usually the transformation of unstructured or semi-structured text or data into a structured form. These methods usually focus on the specific format type, such as log files, json files, xml files, etc.
These works include~\cite{gao2018navigating,du2016spell,klettke2015schema}.
Datamaran~\cite{gao2018navigating} is a tool for extracting structure from semi-structured log datasets, which finds the structures existing in semi-structured log datasets by generalizing log data into several forms such as record patterns and structural templates and divides the log internals by enumerating the possible log endpoints. In addition, there has been a lot of work learning structural properties for data transformation~\cite{Ataccama_ONE,raman2001potter}.

\textbf{Data validation.} Recent efforts on data validation employed 
machine learning techniques, including Google's TensorFlow data 
validation (TFDV) \cite{breck2019data}, Amazon's 
Deequ~\cite{schelter2018automating}, and auto-validation~\cite{song2021auto}. 
TFDV and Deequ allow developers to write declarative data constraints 
to describe how the data should look like. Auto-validation features an 
automatic method to discover patterns from data. However, this method 
requires the assistance of a data lake, and it is difficult to explain 
the relationship between the data lake and the data that needs to be 
verified in actual scenarios. Other notable works in this category are 
\cite{polyzotis2017data,biessmann2021automated}. In addition, 
there are also works on error detection~\cite{abiteboul1995foundations,qahtan2020pattern,berti2018discovery,wang2019uni}, 
aiming at detecting errors in multi-columns of tables, as well as 
methods based on functional dependency\cite{abiteboul1995foundations}) 
and denial constraint\cite{qahtan2020pattern,berti2018discovery}. 
There are also a few data validation 
solutions~\cite{zhang2019sato,yan2018synthesizing,hulsebos2019sherlock} 
besides pattern-based validation approaches. For the detection of rich 
semantics types, Yan et al.~\cite{yan2018synthesizing} used source code 
to generate the type-detection logic, and Satyanarayan et 
al.~\cite{hulsebos2019sherlock} used deep learning methods for semantics 
data type detection. 

\textbf{Data profiling.}
There has been a lot of work on data profiling. abedjan~\cite{abedjan2015profiling}
present a data profiling survey about the relational database. The categories of data profiles include statistical profiles, pattern profiles, embedding method based machine learning and so on. 
Influential works on pattern profiling including
inducing regular expressions from data \cite{gold1978complexity,pitt1993minimum,angluin1987learning} and inducing pattern(domain specific languages) 
from data. Influential works about including patterns from data 
include Potter's Wheel~\cite{raman2001potter}, which leverages the MDL principle 
to rank pattern candidates, and FIDEX~\cite{wang2016fidex}, which allows users to 
provide positive and negative examples and then use these examples to generate patterns 
to find desired data from a dataset. Other notable studies on pattern profiling are
 \cite{fernau2009algorithms,golab2010data}.

 \section{Conclusion}
In this paper, we proposed a self-validated data management system 
for complex nested data. Our system discovers patterns from existing 
data and detects erroneous incoming data. We designed a 
two-step pattern discovery approach by considering both structural 
information and fine-grained semantics of existing configurations. 
To address scalability issues, we proposed optimization techniques 
to incrementally discover patterns. Our experiments demonstrated 
the effectiveness and efficiency of the proposed system as well 
as its superiority over alternative solutions. In addition, the 
experiments on an industrial platform containing thousands of 
applications showed that the data quality rose significantly when 
the database was equipped with our proposed method.
 
\bibliographystyle{abbrv}
\bibliography{reference}

\begin{thebibliography}{10}

\bibitem{kaggle}
Kaggle.
\newblock \url{https://www.kaggle.com/}.

\bibitem{FlashProfile_code}
{FlashProfile}, 2016.
\newblock \url{https://github.com/microsoft/prose}.

\bibitem{xSystem_code}
{xSystem}, 2016.
\newblock \url{https://bitbucket.org/andrewiilyas/xsystem-old/src/}.

\bibitem{Ataccama_ONE}
Ataccama one, 2017.
\newblock \url{https://www.ataccama.com/}.

\bibitem{Deequ}
Deequ, 2018.
\newblock \url{https://github.com/awslabs/deequ}.

\bibitem{TFDV}
Tensorflow data validation, 2019.
\newblock \url{https://www.tensorflow.org/tfx/guide/tfdv}.

\bibitem{Excel}
{Excel}, 2023.
\newblock \url{https://www.microsoft.com/en-us/microsoft-365/excel}.

\bibitem{OpenRefine}
{OpenRefine}, 2023.
\newblock \url{https://openrefine.org/docs}.

\bibitem{PowerBI}
{Power BI: Data Flow}, 2023.
\newblock \url{https://docs.microsoft.com/en-us/power-bi/transform-model/service-dataflows-create-use}.

\bibitem{SSIS}
{SSIS: Data Profiling}, 2023.
\newblock \url{https://docs.microsoft.com/en-us/sql/integration-services/control-flow/data-profiling-task?view=sql-server-ver15}.

\bibitem{Trifacta}
{Trifacta}, 2023.
\newblock \url{https://www.trifacta.com/}.

\bibitem{abedjan2015profiling}
Z.~Abedjan, L.~Golab, and F.~Naumann.
\newblock Profiling relational data: a survey.
\newblock {\em The VLDB Journal}, 24(4):557--581, 2015.

\bibitem{abiteboul1995foundations}
S.~Abiteboul, R.~Hull, and V.~Vianu.
\newblock {\em Foundations of databases}, volume~8.
\newblock Addison-Wesley Reading, 1995.

\bibitem{agichtein2004mining}
E.~Agichtein and V.~Ganti.
\newblock Mining reference tables for automatic text segmentation.
\newblock In {\em KDD}, pages 20--29, 2004.

\bibitem{angluin1987learning}
D.~Angluin.
\newblock Learning regular sets from queries and counterexamples.
\newblock {\em Information and computation}, 75(2):87--106, 1987.

\bibitem{berti2018discovery}
L.~Berti-{\'E}quille, H.~Harmouch, F.~Naumann, N.~Novelli, and T.~Saravanan.
\newblock Discovery of genuine functional dependencies from relational data with missing values.
\newblock {\em PVLDB}, 11(8):880--892, 2018.

\bibitem{biessmann2021automated}
F.~Biessmann, J.~Golebiowski, T.~Rukat, D.~Lange, and P.~Schmidt.
\newblock Automated data validation in machine learning systems.
\newblock {\em Bulletin of the IEEE Computer Society Technical Committee on Data Engineering}, 2021.

\bibitem{Boehme2017}
M.~B{\"{o}}hme, E.~O. Soremekun, S.~Chattopadhyay, E.~Ugherughe, and A.~Zeller.
\newblock Where is the bug and how is it fixed? an experiment with practitioners.
\newblock In E.~Bodden, W.~Sch{\"{a}}fer, A.~van Deursen, and A.~Zisman, editors, {\em Proceedings of the 2017 11th Joint Meeting on Foundations of Software Engineering, {ESEC/FSE} 2017, Paderborn, Germany, September 4-8, 2017}, pages 117--128. {ACM}, 2017.

\bibitem{breck2019data}
E.~Breck, N.~Polyzotis, S.~Roy, S.~Whang, and M.~Zinkevich.
\newblock Data validation for machine learning.
\newblock In {\em MLSys}, 2019.

\bibitem{chu2015tegra}
X.~Chu, Y.~He, K.~Chakrabarti, and K.~Ganjam.
\newblock Tegra: Table extraction by global record alignment.
\newblock In {\em SIGMOD}, pages 1713--1728, 2015.

\bibitem{cortez2011joint}
E.~Cortez, D.~Oliveira, A.~S. da~Silva, E.~S. de~Moura, and A.~H. Laender.
\newblock Joint unsupervised structure discovery and information extraction.
\newblock In {\em Proceedings of the 2011 ACM SIGMOD International Conference on Management of data}, pages 541--552, 2011.

\bibitem{du2016spell}
M.~Du and F.~Li.
\newblock Spell: Streaming parsing of system event logs.
\newblock In {\em ICDM}, pages 859--864, 2016.

\bibitem{elmeleegy2009harvesting}
H.~Elmeleegy, J.~Madhavan, and A.~Halevy.
\newblock Harvesting relational tables from lists on the web.
\newblock {\em PVLDB}, 2(1):1078--1089, 2009.

\bibitem{fernau2009algorithms}
H.~Fernau.
\newblock Algorithms for learning regular expressions from positive data.
\newblock {\em Information and Computation}, 207(4):521--541, 2009.

\bibitem{fisher2005pads}
K.~Fisher and R.~Gruber.
\newblock Pads: a domain-specific language for processing ad hoc data.
\newblock {\em ACM Sigplan Notices}, 40(6):295--304, 2005.

\bibitem{fisher2011pads}
K.~Fisher and D.~Walker.
\newblock The pads project: an overview.
\newblock In {\em Proceedings of the 14th International Conference on Database Theory}, pages 11--17, 2011.

\bibitem{gao2018navigating}
Y.~Gao, S.~Huang, and A.~Parameswaran.
\newblock Navigating the data lake with datamaran: Automatically extracting structure from log datasets.
\newblock In {\em SIGMOD}, pages 943--958, 2018.

\bibitem{golab2010data}
L.~Golab, H.~Karloff, F.~Korn, and D.~Srivastava.
\newblock Data auditor: Exploring data quality and semantics using pattern tableaux.
\newblock {\em PVLDB}, 3(1-2):1641--1644, 2010.

\bibitem{gold1978complexity}
E.~M. Gold.
\newblock Complexity of automaton identification from given data.
\newblock {\em Information and control}, 37(3):302--320, 1978.

\bibitem{Goues2012}
C.~L. Goues, M.~Dewey{-}Vogt, S.~Forrest, and W.~Weimer.
\newblock A systematic study of automated program repair: Fixing 55 out of 105 bugs for {\textdollar}8 each.
\newblock In M.~Glinz, G.~C. Murphy, and M.~Pezz{\`{e}}, editors, {\em 34th International Conference on Software Engineering, {ICSE} 2012, June 2-9, 2012, Zurich, Switzerland}, pages 3--13. {IEEE} Computer Society, 2012.

\bibitem{gulwani2011automating}
S.~Gulwani.
\newblock Automating string processing in spreadsheets using input-output examples.
\newblock {\em ACM Sigplan Notices}, 46(1):317--330, 2011.

\bibitem{Gulzar2021}
M.~A. Gulzar and M.~Kim.
\newblock Optdebug: Fault-inducing operation isolation for dataflow applications.
\newblock In C.~Curino, G.~Koutrika, and R.~Netravali, editors, {\em SoCC '21: {ACM} Symposium on Cloud Computing, Seattle, WA, USA, November 1 - 4, 2021}, pages 359--372. {ACM}, 2021.

\bibitem{hulsebos2019sherlock}
M.~Hulsebos, K.~Hu, M.~Bakker, E.~Zgraggen, A.~Satyanarayan, T.~Kraska, {\c{C}}.~Demiralp, and C.~Hidalgo.
\newblock Sherlock: A deep learning approach to semantic data type detection.
\newblock In {\em KDD}, pages 1500--1508, 2019.

\bibitem{ilyas2018extracting}
A.~Ilyas, J.~M. da~Trindade, R.~C. Fernandez, and S.~Madden.
\newblock Extracting syntactical patterns from databases.
\newblock In {\em ICDE}, pages 41--52, 2018.

\bibitem{Kirschner2020}
L.~Kirschner, E.~O. Soremekun, and A.~Zeller.
\newblock Debugging inputs.
\newblock In G.~Rothermel and D.~Bae, editors, {\em {ICSE} '20: 42nd International Conference on Software Engineering, Seoul, South Korea, 27 June - 19 July, 2020}, pages 75--86. {ACM}, 2020.

\bibitem{klettke2015schema}
M.~Klettke, U.~St{\"o}rl, and S.~Scherzinger.
\newblock Schema extraction and structural outlier detection for json-based nosql data stores.
\newblock {\em Datenbanksysteme f{\"u}r Business, Technologie und Web (BTW 2015)}, 2015.

\bibitem{kozielski2009new}
S.~Kozielski and R.~Wrembel.
\newblock {\em New Trends in Data Warehousing and Data Analysis}.
\newblock Springer, 2009.

\bibitem{padhi2018flashprofile}
S.~Padhi, P.~Jain, D.~Perelman, O.~Polozov, S.~Gulwani, and T.~Millstein.
\newblock Flashprofile: a framework for synthesizing data profiles.
\newblock {\em Proceedings of the ACM on Programming Languages}, 2(OOPSLA):1--28, 2018.

\bibitem{pitt1993minimum}
L.~Pitt and M.~K. Warmuth.
\newblock The minimum consistent dfa problem cannot be approximated within any polynomial.
\newblock {\em Journal of the ACM}, 40(1):95--142, 1993.

\bibitem{polyzotis2017data}
N.~Polyzotis, S.~Roy, S.~E. Whang, and M.~Zinkevich.
\newblock Data management challenges in production machine learning.
\newblock In {\em SIGMOD}, pages 1723--1726, 2017.

\bibitem{qahtan2020pattern}
A.~Qahtan, N.~Tang, M.~Ouzzani, Y.~Cao, and M.~Stonebraker.
\newblock Pattern functional dependencies for data cleaning.
\newblock {\em PVLDB}, 13(5):684--697, 2020.

\bibitem{raman2001potter}
V.~Raman and J.~M. Hellerstein.
\newblock Potter's wheel: An interactive data cleaning system.
\newblock In {\em VLDB}, pages 381--390, 2001.

\bibitem{schelter2019unit}
S.~Schelter, F.~Biessmann, D.~Lange, T.~Rukat, P.~Schmidt, S.~Seufert, P.~Brunelle, and A.~Taptunov.
\newblock Unit testing data with deequ.
\newblock In {\em SIGMOD}, pages 1993--1996, 2019.

\bibitem{schelter2018automating}
S.~Schelter, D.~Lange, P.~Schmidt, M.~Celikel, F.~Biessmann, and A.~Grafberger.
\newblock Automating large-scale data quality verification.
\newblock {\em PVLDB}, 11(12):1781--1794, 2018.

\bibitem{song2021auto}
J.~Song and Y.~He.
\newblock Auto-validate: Unsupervised data validation using data-domain patterns inferred from data lakes.
\newblock In {\em SIGMOD}, pages 1678--1691, 2021.

\bibitem{Teoh2019}
J.~Teoh, M.~A. Gulzar, G.~H. Xu, and M.~Kim.
\newblock Perfdebug: Performance debugging of computation skew in dataflow systems.
\newblock In {\em Proceedings of the {ACM} Symposium on Cloud Computing, SoCC 2019, Santa Cruz, CA, USA, November 20-23, 2019}, pages 465--476. {ACM}, 2019.

\bibitem{wang2019uni}
P.~Wang and Y.~He.
\newblock Uni-detect: A unified approach to automated error detection in tables.
\newblock In {\em SIGMOD}, pages 811--828, 2019.

\bibitem{wang2016fidex}
X.~Wang, S.~Gulwani, and R.~Singh.
\newblock {FIDEX}: filtering spreadsheet data using examples.
\newblock {\em ACM SIGPLAN Notices}, 51(10):195--213, 2016.

\bibitem{DBLP:journals/jss/WangCZW22}
Z.~Wang, T.~P. Chen, H.~Zhang, and S.~Wang.
\newblock An empirical study on the challenges that developers encounter when developing apache spark applications.
\newblock {\em J. Syst. Softw.}, 194:111488, 2022.

\bibitem{Wen2018}
M.~Wen, J.~Chen, R.~Wu, D.~Hao, and S.~Cheung.
\newblock Context-aware patch generation for better automated program repair.
\newblock In M.~Chaudron, I.~Crnkovic, M.~Chechik, and M.~Harman, editors, {\em Proceedings of the 40th International Conference on Software Engineering, {ICSE} 2018, Gothenburg, Sweden, May 27 - June 03, 2018}, pages 1--11. {ACM}, 2018.

\bibitem{Wong2016}
W.~E. Wong, R.~Gao, Y.~Li, R.~Abreu, and F.~Wotawa.
\newblock A survey on software fault localization.
\newblock {\em {IEEE} Trans. Software Eng.}, 42(8):707--740, 2016.

\bibitem{yan2018synthesizing}
C.~Yan and Y.~He.
\newblock Synthesizing type-detection logic for rich semantic data types using open-source code.
\newblock In {\em SIGMOD}, pages 35--50, 2018.

\bibitem{Zeller2002}
A.~Zeller and R.~Hildebrandt.
\newblock Simplifying and isolating failure-inducing input.
\newblock {\em {IEEE} Trans. Software Eng.}, 28(2):183--200, 2002.

\bibitem{zhang2019sato}
D.~Zhang, Y.~Suhara, J.~Li, M.~Hulsebos, {\c{C}}.~Demiralp, and W.-C. Tan.
\newblock Sato: Contextual semantic type detection in tables.
\newblock {\em PVLDB}, 13(11):1835--1848, 2020.

\bibitem{Zhang2020}
Q.~Zhang, J.~Wang, M.~A. Gulzar, R.~Padhye, and M.~Kim.
\newblock Bigfuzz: Efficient fuzz testing for data analytics using framework abstraction.
\newblock In {\em 35th {IEEE/ACM} International Conference on Automated Software Engineering, {ASE} 2020, Melbourne, Australia, September 21-25, 2020}, pages 722--733. {IEEE}, 2020.

\end{thebibliography}

\end{document}